\documentclass[aps,prb,reprint,superscriptaddress,showpacs,longbibliography,floatfix,footnoteinbib]{revtex4-2}
\usepackage{amsmath,amssymb,graphicx,units}
\usepackage[plainpages=false,pdfpagelabels,colorlinks=true,linkcolor=red,urlcolor=blue,citecolor=blue,pdftitle={Title},pdfauthor={},pdfdisplaydoctitle=true,pdfduplex=DuplexFlipLongEdge]{hyperref}
\usepackage{multirow}
\usepackage{setspace}
\usepackage{bm}
\usepackage{lipsum} 

\begin{document}

\title{Probing phase transitions with correlations in configuration space}

\author{Wen-Yu Su}
\affiliation{Key Laboratory of Quantum Theory and Applications of MoE, Lanzhou Center for Theoretical Physics, and Key Laboratory of Theoretical Physics of Gansu Province, Lanzhou University, Lanzhou, Gansu 730000, China}

\author{Yu-Jing Liu}
\affiliation{Key Laboratory of Quantum Theory and Applications of MoE, Lanzhou Center for Theoretical Physics, and Key Laboratory of Theoretical Physics of Gansu Province, Lanzhou University, Lanzhou, Gansu 730000, China}

\author{Nvsen Ma}
\email{nvsenma@buaa.edu.cn}
\affiliation{School of Physics, Beihang University, Beijing 100191, China}

\author{Chen Cheng}
\email{chengchen@lzu.edu.cn}
\affiliation{Key Laboratory of Quantum Theory and Applications of MoE, Lanzhou Center for Theoretical Physics, and Key Laboratory of Theoretical Physics of Gansu Province, Lanzhou University, Lanzhou, Gansu 730000, China}

\begin{abstract}

In principle, the probability of configurations, determined by the system's partition function or wave function, encapsulates essential information about phases and phase transitions. Despite the exponentially large configuration space, we show that the generic correlation of distances between configurations, with a degree of freedom proportional to the lattice size, can probe phase transitions using importance sampling procedures like Monte Carlo simulations. The distribution of sampled distances varies significantly across different phases, suggesting universal critical behavior for uncertainty and participation entropy. For various classical spin models with different phases and transitions, finite-size analysis based on these quantities accurately identifies phase transitions and critical points. Notably, in all cases, the critical exponent derived from the uncertainty of distances equals the anomalous dimension governing real-space correlation decay. Thus, configuration space correlations, defined by distance uncertainties, share the same decay ratio as real-space correlations, determining the universality class of phase transitions. This work applies to diverse lattice models with different local degrees of freedom, e.g., two levels for Ising-like models, discrete multi-levels for $q$-state clock models, and continuous local levels for the $XY$ model, offering a robust, alternative method for understanding complex phases and transitions.

\end{abstract}
\maketitle

\section{Introduction}
\label{sec:introduction}

Phase transitions are fundamental in statistical and condensed matter physics~\cite{Landau1999, Sachdev_2011}. According to the conventional Landau-Ginzburg-Wilson (LGW) paradigm, the critical phenomena can be described by the fluctuations of an order parameter that distinguishes two phases separated by the critical point~\cite{Landau1999,wilson}, close to which the correlation length of the model diverges to infinity. The phase transition can be understood in field theory through the long-wavelength limit, where the behaviors of physical quantities, including the order parameter, are governed by corresponding universal critical exponents~\cite{cardy}. Nowadays, calculating and analyzing the order parameters and critical exponents with numerical approaches, such as Monte Carlo (MC) simulation or tensor network algorithms, is a standard procedure to understand the critical phenomenon, which has been successful in describing continuous phase transitions and Berezinskii-Kosterlitz-Thouless (BKT) transitions in various lattice models~\cite{Sandvik2010,crisix1, Pascal,zhenggu, Challa1986, Tomita2002, Hwang2009, Baek2010, Kumano2013, Li2022_qstateclock}. 

In practical applications, when facing complex models or complicated phase transitions, it is not always straightforward to define a proper order parameter that can encapsulate the difference between phases~\cite{guangming, KZhao}. While people keep trying to dig out new quantities that can exhibit singularity or non-monotonic behavior in tracing the phase transition, from the spin stiffness in the BKT transition~\cite{fisher1}, the eigen microstate in statistical ensemble~\cite{Hu2019} to the newly proposed average sign of sampling weight in quantum MC simulations with a sign problem~\cite{sign1,nvsen2024}, those quantities are usually applicable in the specific systems or phase transitions. As the phase transition manifests universal criticality, one would wonder whether one can find some universal procedures analyzing or providing preliminary hints of the phase transition in generic lattice systems with less prior knowledge. 

The recently developed approaches of integrating machine learning (ML) ideas with the Monte Carlo simulations have proposed an alternative way to study phase transitions, that is, handling the MC-obtained results as a generic data collection without taking account of their physical insights~\cite{Carrasquilla2017, Zhang2019, Ding2022}. While processing the data collection of physical quantities with different ML-related methods has successfully captured the phase transition in various spin systems~\cite{Suchsland2018, ising3, Beach2018, xy3, Funai2020, Shiina2020, Giataganas_2022}, some recent studies put more effort into procedures that can quantify the raw data of spin configurations with scaling behaviors~\cite{Miyajima2021, Miyajima2023, santos1, Santos2021prxq,arnoldprx,arnoldprr}. The latter approach, which in principle does not rely on traditional physical quantities in specific systems, has the potential to generate new quantities that behave like order parameters in a more generic phase-transition analyzing procedure. 

On the other hand, while we mainly focus on physical quantities in real or momentum space, measurement in Hilbert space has recently been widely used in the dynamic phase transitions of close quantum systems. Especially for localization problems, the extended or localized state can be distinguished by the participation entropy defined from the probability of the wave function projected to the chosen basis~\cite{Backer2019, Sutradhar2022, Yao2023, Mace2019,  Cheng2023}. For the localization transition in quantum many-body systems, the structure of the wave function is not necessarily related to the specific system.

Inspired by ML-related ideas and the wave function-based analysis in localization problems, we consider the possibility of probing phase transitions with a more generic procedure in the Hilbert space, which is the same for various lattice models that share the same local degree of freedom. Due to the exponential dimension of the Hilbert space that is unachievable in practical numerical approaches, we instead look at the bounded correlation in the configuration space and combine the Monte Carlo method to generate sampled configurations. More specifically, the phase transitions are characterized by the very basic data analysis on the data set of distances between sampled configurations, and the related functions of distances can be considered as generic correlations in configuration space. When applied to several classical spin models with different types of phase transitions, i.e., the first order, continuous, and BKT phase transitions, our proposal successfully catches all the critical points. Moreover, the numerical results suggest that the correlations in the configuration space exhibit universal critical behaviors, and the related critical exponent can be connected to the correlations in the real space.   

The rest of the paper is organized as follows. In Sec.~\ref{sec:basic_idea}, we introduce the key conception of the work, i.e., the distances between sampled configurations, and the probability, uncertainty, and participation entropy extracted from it, and discuss how these measurements can probe the phase transition with the example of the Ising model. In Sec.~\ref{sec:XY_and_q_state_model}, we extend our proposal in the $q$-state clock models for the integer $q$ local levels and the $XY$ limit with continuous distances. Sec.~\ref{sec:critical_exponent} further discusses the universal criticality of the uncertainty and demonstrates that the critical exponent of the uncertainty equals the anomalous dimension, which characterizes the universality class of the phase transition. The summary and discussion are made in Sec.~\ref{sec:sum}. Moreover, technical details can be found in the Appendix sections. 

\section{Phase transition in configuration space}
\label{sec:basic_idea}

The standard way of investigating phases and phase transitions relies on the expectation values of the order parameter or other related physical quantities, such as the specific heat and correlations in real space. In a statistical system described by the Hamiltonian $\hat{H}$, no matter quantum or classical, a generic observable, i.e., the expectation value of a generic operator $\hat{O}$, is estimated by
\begin{align}
    \langle \hat{O} \rangle = \frac{\mathrm{Tr} {e^{-\beta \hat{H}} \hat{O}}}{Z} 
    \label{eq:partition_function},
\end{align}
where $Z=\mathrm{Tr} e^{-\beta \hat{H}}$ is the partition function at temperature $ T$ with $\beta=1/T$. However, for generic lattice models, the problem with the Hamiltonian matrix living in an exponentially large Hilbert space can not be exactly solved and relies on numerical approaches. Among them, the Monte Carlo simulation performs the importance sampling procedure and generates a finite number ${{\cal N}_s}$ of configurations $\{s\}$ which obeys the probability $p(s)\equiv e^{-\beta E_s}/\sum_{s} e^{-\beta E_s}$ with $E_s$ the energy of $s$. Then, a good approximation of physical quantities can be obtained as
\begin{align}
    \langle \hat{O} \rangle = \sum_{s} p(s) O_s \approx  \frac{1}{{\cal N}_s}\sum_{s}^{{\cal N}_s} O_s,
    \label{eq:Ps}
\end{align}
and the analysis of phases and phase transitions is carried out based on the obtained physical quantities. For a closed quantum system described by the wavefunction $|\Psi\rangle =\sum_s\psi_s|s\rangle$, one sometimes follows the similar way probing phase transitions, where $p(s)\equiv |\psi_s|^2$ and the sampling procedure can be realized by numerical MC simulations or the quantum collapse measuring in experiments~\cite{Guo2020, Guo2021, Yao2023}.  

The present work raises and aims to answer the following question: whether and how can we probe phase transitions from $p(s)$, or in practice, a sampled data collection of configurations $\{ s \}$, without the traditional order parameter or other widely used physical quantities?

\begin{figure}[!t]
\includegraphics[width=\columnwidth]{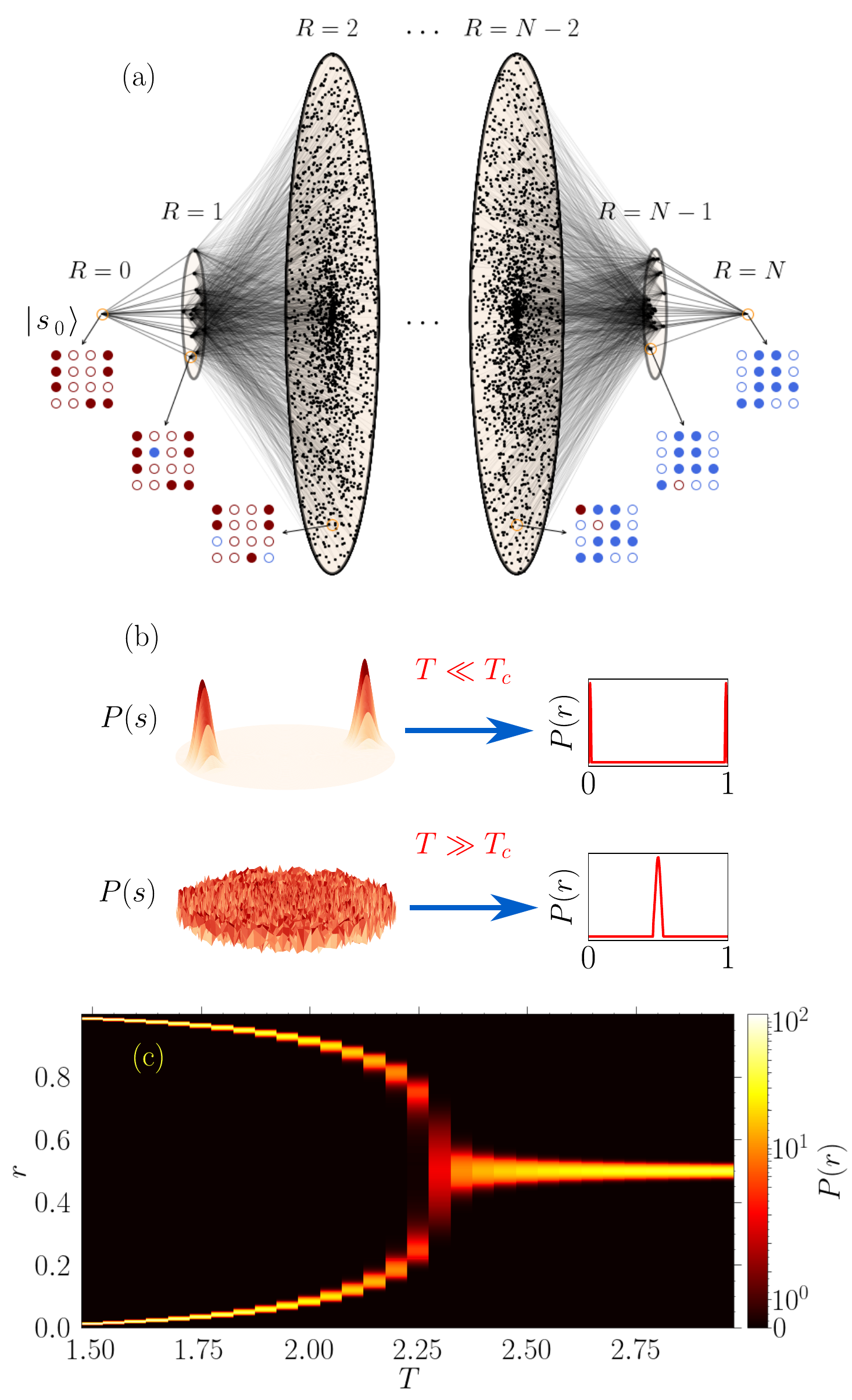}
\caption{(a) A schematic plot of the configuration space for the lattice model with two-level local sites. The filled/empty circle demonstrates on-site state $|1\rangle$/$|0\rangle$ on a square lattice, and the color is to guide the eye. The configurations are arranged by the distance from an arbitrary configuration $|s_0\rangle$ at the leftmost side. (b) The different $P(s)$ in the configuration space leads to the distinct distribution of distances $P(r)$. For the Ising model, two peaks of $P(s)$ at low temperatures correspond to thermal fluctuations around two degenerate ground states. (c) The distribution $P(r)$ as a function of temperature indicates the two phases and a phase transition, where the data is from Monte Carlo simulations on a square lattice with the system length $L=80$.}
\label{fig:schematic}
\end{figure}

\subsection{Distances between configurations}

It is worth mentioning that $p(s)$ is commonly used for localization problems in closed quantum systems, where the participation ratio and participation entropy based on the probability $p(s)$ of the eigenstate manifest the dynamical localization transition for systems with relatively small Hilbert space~\cite{Visscher1972,Mace2019,Cheng2023,Guo2020}. However, in the sampling procedures for large system sizes, the exact $p(s)$, as well as the participation ratio or entropy, is unachievable.

To avoid the exponentially large degree of freedom of $p(s)$, our proposal relies on the correlations between configurations, which are specifically quantified as functions of the 1-norm distance in the configuration space. For a lattice model with two energy levels on a local site, the normalized distance between two configurations is defined as
\begin{align}
    r_{\alpha,\beta} = R_{\alpha,\beta}/N = \sum_i |s^\alpha_i - s^\beta_i|/N,
    \label{eq:distance_q2}
\end{align}
where $|s^\alpha\rangle=|s^\alpha_0,s^\alpha_1,\cdots,s^\alpha_{N-1}\rangle$ is a configuration of system with $N$ sites and $s^\alpha_i\in\{0,1\}$ on site $i$. Obviously, $r$ has $N+1$ discrete values in the range $[0,1]$ with $R=0,1,\cdots,N$. In other words, the degree of freedom for distances is proportional to $N$ instead of $2^N$. As displayed in Fig.~\ref{fig:schematic}(a), starting from an arbitrary configuration $|s_0\rangle$ on the left, the configuration space can be grouped as configuration layers according to the distance to $|s_0\rangle$. The population of a layer with a certain distance can be easily solved as a simple combination problem, and the probability of a random configuration that has a distance $r$ to $|s_0\rangle$ is
\begin{align}
    p(r) = \binom NR /{2^N},
    \label{eq:Pr_inf_T}
\end{align}
which approximates a normal distribution for large $N$ (see Appendix~\ref{appendix:Pr_analysis}). For generic systems, this equation also gives the probability of distances between all possible pairs of configurations in the infinite-temperature limit, where all configurations are uniformly distributed. 

While the low temperature $p(s)$ is system-dependent, we first take the simplest Ising model as an example, which reads 
\begin{align}
    H_\mathrm{Ising} = -J\sum_{\langle ij\rangle} \sigma^z_i \sigma^z_j,
    \label{eq:H_Ising}
\end{align}
where the sum $\langle ij \rangle$ runs over the nearest neighbor sites on a square lattice and $J>0$ indicates the ferromagnetic interactions. At zero temperature, the only nonzero probability of the configuration appears evenly on the two degenerated ground states $|000\cdots\rangle$ and $|111\cdots\rangle$, which further leads to $p(r=0)=p(r=1)=0.5$. As displayed in the schematic plot in Fig.~\ref{fig:schematic}(b), the distribution of distances between configurations can apparently distinguish the low-temperature ordered and the high-temperature disordered states, at least for the Ising model. Here and after, the capital $P$ denotes the probability distribution, which is more convenient when comparing different system sizes. Moreover, in the MC sampled procedure with a finite number of configurations, the numerically obtained $P(r)$ as a function of $T$ clearly demonstrates two phases and a possible phase transition at a finite system size, as displayed in Fig.~\ref{fig:schematic}(c).

\subsection{Phase transition of the Ising model}

In order to get more information about the phase transition, such as the critical point and critical exponents in the thermodynamic limit, one needs quantities related to $p(r)$ with scaling behaviors. Considering $\{r\}$ as a generic data set, the simplest analysis is based on the average value and the uncertainty (the standard deviation) of distances 
\begin{align}
    \overline{r} =& \frac{1}{{\cal N}_r}\sum_{m=1}^{{\cal N}_r} r_m, \\
    \sigma_r =& \sqrt{\frac{(r_m - \overline{r})^2}{{\cal N}_r}},
    \label{eq:r_mean_sigma}
\end{align}
with $r_m\in\{r\}$ and $\{r\}$ is the data collection from the Monte Carlo samplings and ${\cal N}_r$ is the total number of distances (see Appendix~\ref{appedex:sampling_details} for details of sampled configurations and distances).

\begin{figure}[!t]
\includegraphics[width=\columnwidth]{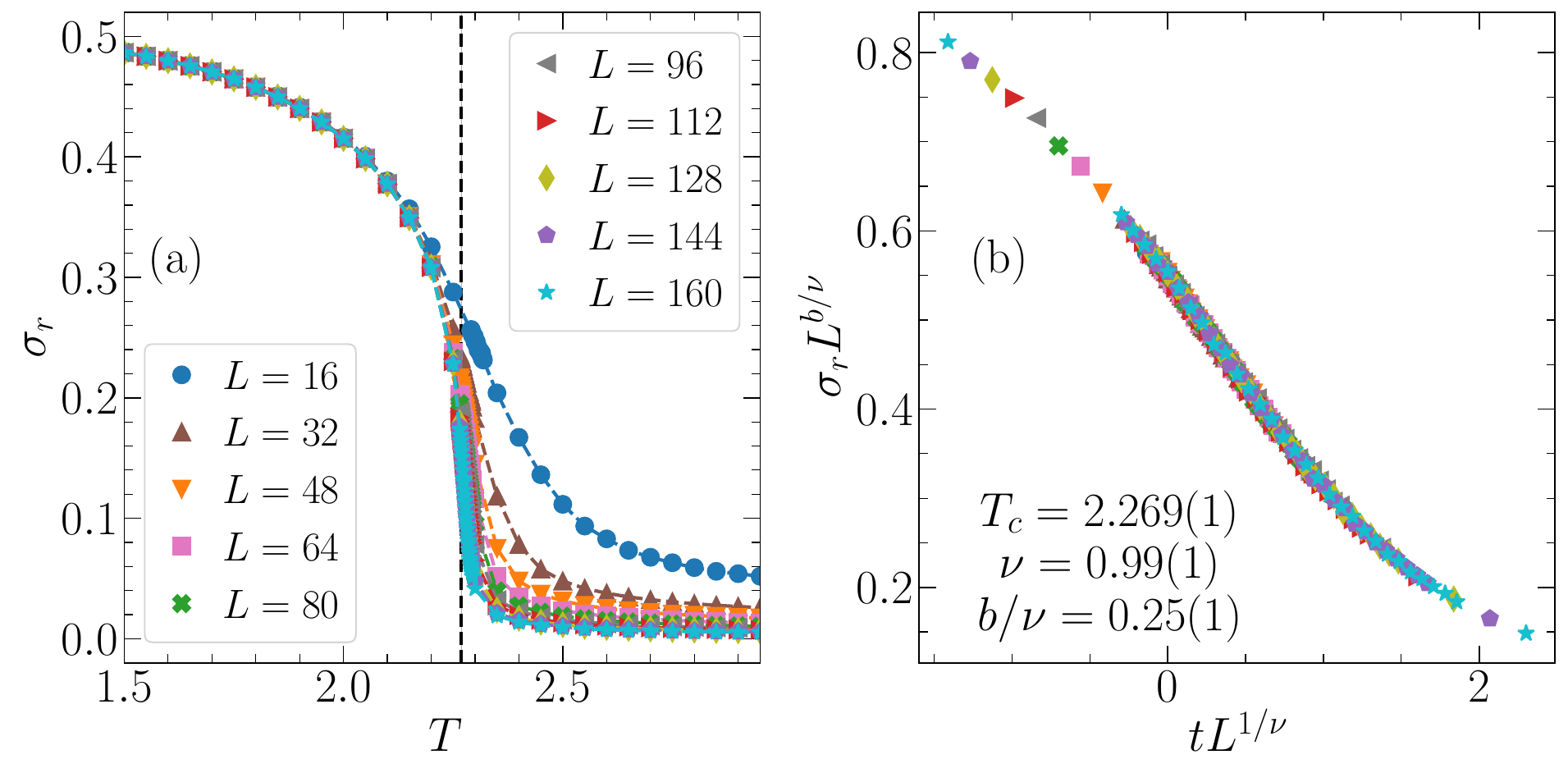}
\caption{(a) The standard deviation $\sigma_r$ with respect to temperature $T$ for the Ising model on square lattices with different lattice lengths. The vertical dashed line marks the critical point from the finite-size scaling of $\sigma_r$, with the shaded area depicting the error.  (b) The best data collapse of the scaling with the obtained critical point and exponents. }
\label{fig:sigma_q2}
\end{figure}

Under an unbiased sampling procedure, one has $\overline{r}=0.5$ for the Ising model at all temperatures due to the spin inversion symmetry. The uncertainty of distances is also accessible in the two extremal cases: At zero temperature, $\sigma_r=0.5$ for two-fold degenerate ground state despite the system size; At the limit $T\rightarrow\infty$, $p(r)$ approximates a normal distribution with $\sigma_r\propto 1/L$. As displayed in Fig.~\ref{fig:sigma_q2}(a), the numerically obtained $\sigma_r$ decreases as the temperature increases, and the values agree well with the above analysis in the low/high-temperature limit. Besides, the different size dependencies of $\sigma_r$ are found in different phases, from which we can reasonably assume that $\sigma_r$ exhibits a scaling behavior close to the phase transition point with the following form 
\begin{align}
    \sigma_r\sim |t|^{b}
    \label{eq:scaling_second},
\end{align}
where $t=|T-T_c|/T$ with $T_c$ the critical temperature and $b$ is the scaling exponent related to $\sigma_r$. Thus, using the finite-size scaling hypothesis with the continuous phase transition in the Ising model, we can get the size dependence of $\sigma_r$ as 
\begin{align}
    \sigma_r L^{b/\nu} = {\cal F}_\mathrm{con}(tL^{1/\nu}),
    \label{eq:scaling_con}
\end{align}
with $\nu$ the correlation length exponent. In this way, one can plot $\sigma_rL^{b/\nu}$ versus $tL^{1/\nu}$ for different sizes, and all the data should collapse to the same curve with the correct critical information. 

In the practical scaling procedure, we assume no foreknowledge of the critical point or exponents and numerically search for the best data collapse in the three-dimensional parameter space $\{T_c,\nu,b\}$  (see Appendix~\ref{appedix:scaling_details} for scaling details). As shown in Fig.~\ref{fig:sigma_q2}(b), a nice data collapse is obtained with all data points from different system sizes collapsing to a smooth curve. The numerically obtained $T_c=2.269(1)$ agrees with the analytical result, and the correlation length exponent $\nu=0.99(1)$ also matches the true value $\nu=1$~\cite{Sandvik2010}. 

\begin{figure}[!t]
\includegraphics[width=\columnwidth]{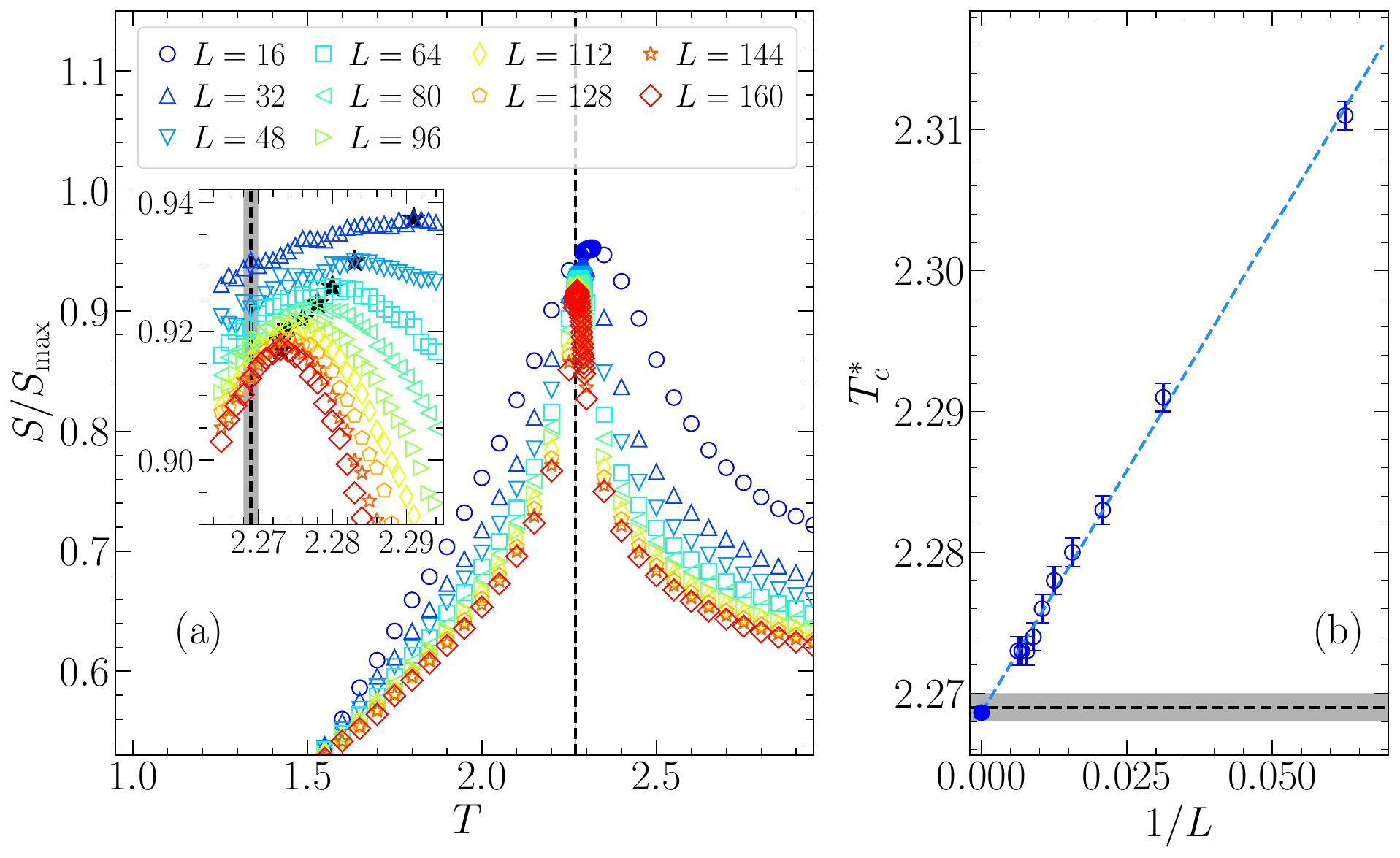}
\caption{(a) The participation entropy $S/S_\mathrm{max}$ verses $T$ for the Ising model on square lattices with  different $L$. The inset shows the zoom-in close to the transition point, and the temperature at the maximum of $S$ indicates the finite size critical point $T_c^*$. (b) The finite size extrapolation of $T_c^*$. The error bar of $T_c^*(L)$ is determined as the interval of simulated temperatures, and the error bar of $1/L\rightarrow 0$ depicts the uncertainty from the fitting. The vertical/horizontal dashed line in (a)/(b) marks the critical point from the finite-size scaling of $\sigma_r$, with the shaded area depicting the error.}
\label{fig:S_q2}
\end{figure}

In addition, one can always define the participation entropy or the Shannon entropy $S$ from the probability $p(r)$ as 
\begin{align}
    S_r = -\sum_r p(r) \log p(r).
    \label{eq:entropy}
\end{align}
Different from $p(s)$, which has exponentially large degrees of freedom and is inaccessible in MC simulations, $p(r)$ is complete even in the sampling procedure with a small fraction of configurations. For the Ising model or other lattice models with two local levels, $r$ is discrete with $N+1$ possible values and one has $S_r\in[0, S_\mathrm{max}]$ with $S_\mathrm{max} = \log (N+1)$. The numerically obtained $S_r$ is displayed in Fig.~\ref{fig:S_q2}(a), where a sharp peak emerges at the finite-size critical point $T_c^*(L)$ for each system size, and its position gets closer to the critical point in the thermodynamic limit as the system size increases. In Fig.~\ref{fig:S_q2}(b), we perform a finite-size extrapolation with $T_c^*(L)-T_c^*(\infty)\sim 1/L$ following the continuous phase transition, and the obtained $T_c^*(\infty)$ agrees well with the finite-size scaling result. 

\subsection{Determine the phase transition type}

\begin{figure}[!t]
\includegraphics[width=0.8\columnwidth]{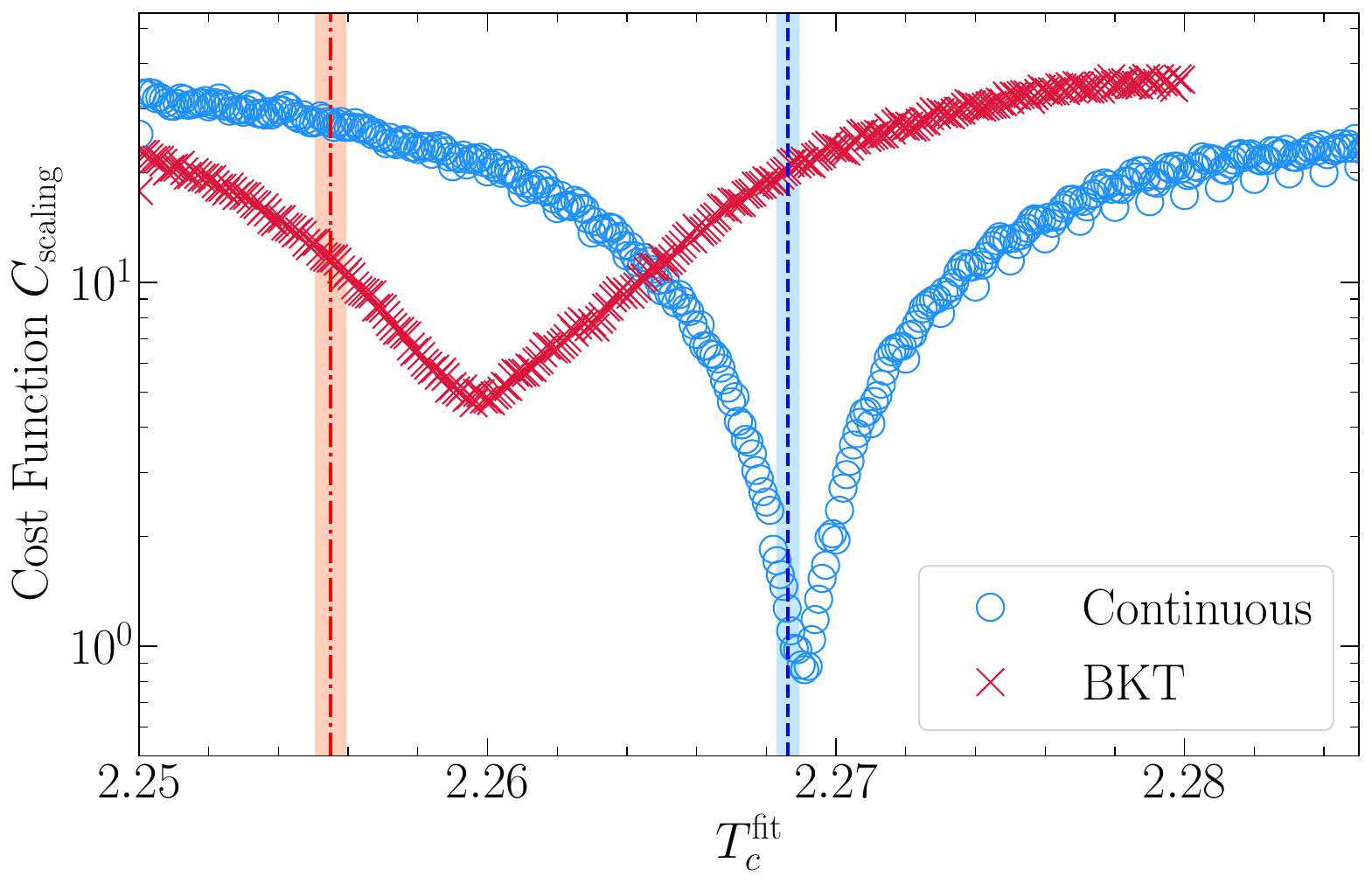}
\caption{The cost functions $C_{\mathrm{scaling}}$ [see Eq.~\eqref{eq:cost_function} and Appendix~\ref{appedix:scaling_details}] in scaling procedures of $\sigma_r$ as a function of $T_c^\mathrm{fit}$. The vertical dashed line depicts $T_c^*$ from the finite size extrapolation of $S_r$, with the shading area depicting the error. The blue and red markers/lines depict procedures adopting the continuous and BKT type of phase transitions, respectively.}
\label{fig:transition_type_q2}
\end{figure}

The above analysis adopts the known continuous phase transition of the Ising model. Even without this knowledge, one can numerically judge whether a certain transition type is correct by checking the matching between the scaling results of $\sigma_r$ and the critical extrapolation of $S_r$. As shown in Fig.~\ref{fig:transition_type_q2}, the best $T_c^\mathrm{fit}$, which is located where the cost function $C_{\mathrm{scaling}}$ in the scaling procedure of $\sigma_r$ reaches minimum, is close to $T_c^*$ from the extrapolation of $S_r$. Here, the smaller $C_{\mathrm{scaling}}$ indicates a better data collapse, and the best $[b,\nu]$ is fixed here to show the $T_c^\mathrm{fit}$ dependence of $C_{\mathrm{scaling}}$. At the same time, the obtained $\nu\approx 1$ in Fig.~\ref{fig:sigma_q2}(b) can rule out the first-order phase transition, which has $\nu=1/d=0.5$ in a two-dimensional spin system~\cite{fisher1982}. 

We also check the possibility of a BKT phase transition~\cite{Kosterlitz1973,Kosterlitz1974}, where the correlation length $\xi$ diverges as
\begin{align}
   \xi\sim e^{c/\sqrt{t}}
    \label{eq:BKT_crrl}
\end{align}
with a system-dependent variable $c$. Assuming that $\sigma_r$ exhibits power-law behavior in BKT transition as $\sigma_r\sim \xi^{-b}$, one can write the following scaling form as
\begin{align}
    \sigma_r L^{b}={\cal F}_\mathrm{BKT}(L/e^{c/\sqrt{t}}),
    \label{eq:scaling_BKT}
\end{align}
with $b$ the critical exponent of $\sigma_r$. Performing the same scaling procedure for the BKT phase transition following Eq.~\eqref{eq:scaling_BKT}, the minimum cost function in Fig.~\ref{fig:transition_type_q2} is much larger than the continuous fitting. Meanwhile, the critical temperature obtained from extrapolating following the BKT form $T_c^*(L)-T_c^*(\infty)\sim 1/\log^2 L$ is far away from the best $T_c^\mathrm{fit}$ with least $C_{\mathrm{scaling}}$. Therefore, the numerical experiment prefers a continuous phase transition in the Ising model. 

Thus, the uncertainty $\sigma_r$ and the participation entropy $S_r$ of the distances between the sampled configurations can give us the accurate critical point and the correlation length exponent for the Ising model even assuming no foreknowledge about the phase and phase transition. We left the discussions on the critical exponent $b$ for $\sigma_r$ in later sections. 

\section{The \texorpdfstring{$q$}{Lg}-state clock models}
\label{sec:XY_and_q_state_model}

In this section, we carry out a similar procedure using the correlations between sampled spin configurations to study the critical behavior in the $q$-state clock model~\cite{Elitzur1979,Cardy_1980,Tobochnik1982,Challa1986,Chatterjee2018,Li2020,youjin2022} with Hamiltonian
\begin{align}
    H = -J \sum_{\langle ij\rangle} \vec{S}_i \cdot \vec{S}_j = -J \sum_{\langle ij\rangle} \cos(\theta_i-\theta_j),
    \label{eq:ham_qstate}
\end{align}
where $J$ is positive with the sum $\langle ij\rangle$ runs over the nearest neighbor sites. The classical spin $\vec{S}_i=(\cos(\theta_i),\sin(\theta_i))$ lives on site $i$ has discrete $\theta=2\pi\gamma/q$ with integer $\gamma \in \{0, 1,\cdots,q - 1 \}$. On the two-dimensional square lattice, the $q$-state models contain rich phases and phase transitions~\cite{Elitzur1979}. For $2\leq q\leq 4$, the model features a continuous phase transition separating the magnetically ordered and disordered phases as temperature increases. For $q\geq 5$, an intermediate quasi-long-range-ordered phase emerges between the ordered and disordered phase, and the model undergoes two BKT phase transitions~\cite{Elitzur1979}. When $q\rightarrow\infty$, the $q$-state clock model becomes the $XY$ model with continuously varying $\theta$ in $[0,2\pi)$, where only one BKT phase transition from the quasi-long-range-ordered critical phase to the high-temperature disorder phase left.

While the $2$-state clock mode equals the Ising model, the $q>2$ cases contain more than two levels on a local site. Nevertheless, we can still define the normalized 1-norm distance as 
\begin{align}
    r_{\alpha,\beta} = r(\vec{s}_\alpha,\vec{s}_\beta) = \sum_i r(\theta_i^\alpha,\theta_i^\beta)/(N\pi),
    \label{eq:distance_qstate}
\end{align}
where $\vec{s}_\alpha=\{\theta_0^\alpha,\theta_1^\alpha,\cdots,\theta_i^\alpha,\cdots\}$ stores the configuration labeled by $\alpha$. Here $r(\theta_i^\alpha,\theta_i^\beta) = \min \{|\theta_i^\alpha-\theta_i^\beta|, 2\pi - |\theta_i^\alpha-\theta_i^\beta|\}$ is restricted within $[0,\pi]$ and the two-state distance $r_{\alpha,\beta}$ is restricted to $[0,1]$. It is easy to find that Eq.~\eqref{eq:distance_qstate} degenerates to Eq.~\eqref{eq:distance_q2} of the Ising model at $q=2$.

\begin{figure}[!t]
\includegraphics[width=\columnwidth]{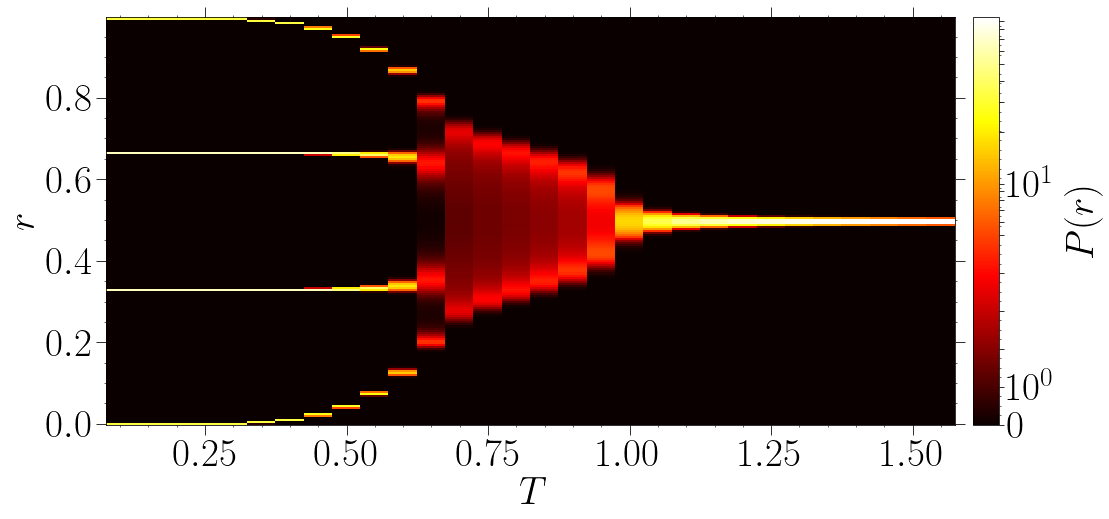}
\caption{The distribution $P(r)$ as a function of temperature for the $6$-clock state model on a square lattice with the system length $L=174$.}
\label{fig:Pr_pcolor_q6}
\end{figure}

\subsection{6-state clock model}

We start with the $q=6$ case to study the two successively BKT phase transitions, which are completely different from the critical behaviors of the Ising model. Nevertheless, we can perform MC importance samplings and obtain a data set of distances $\{r\}$ between the sampled configurations. The 
corresponding distribution $P(r)$ on a finite lattice displayed in Fig.~\ref{fig:Pr_pcolor_q6} clearly manifests three different regions and two possible phase transitions. It is known that the $6$-state clock model has a six-fold degenerate ground state at low temperatures, where all spins point to the same angle  $\theta$ and $\theta=2\pi\gamma/6$ with $\gamma \in \{0, 1,\cdots,5 \}$. These six ground states have the same weight, and there are only four possible values of the distances between any two of them as $r=\{0,1/3,2/3,1\}$~\cite{Su2023}. At high temperatures, the $6$-state clock model becomes completely disordered with a uniform distribution of all the configurations, leading to a normally distributed $P(r)$ with the expectation value $\overline{r}=0.5$. The numerically obtained $P(r)$ shown in Fig.~\ref{fig:Pr_pcolor_q6} at low and high temperatures agrees well with the above analysis. Besides, an intermediate region with a spread $P(r)$ different from either of those two extreme situations is found at intermediate temperatures, corresponding to the critical region with a quasi-long-range order. 

\begin{figure}[!t]
\includegraphics[width=\columnwidth]{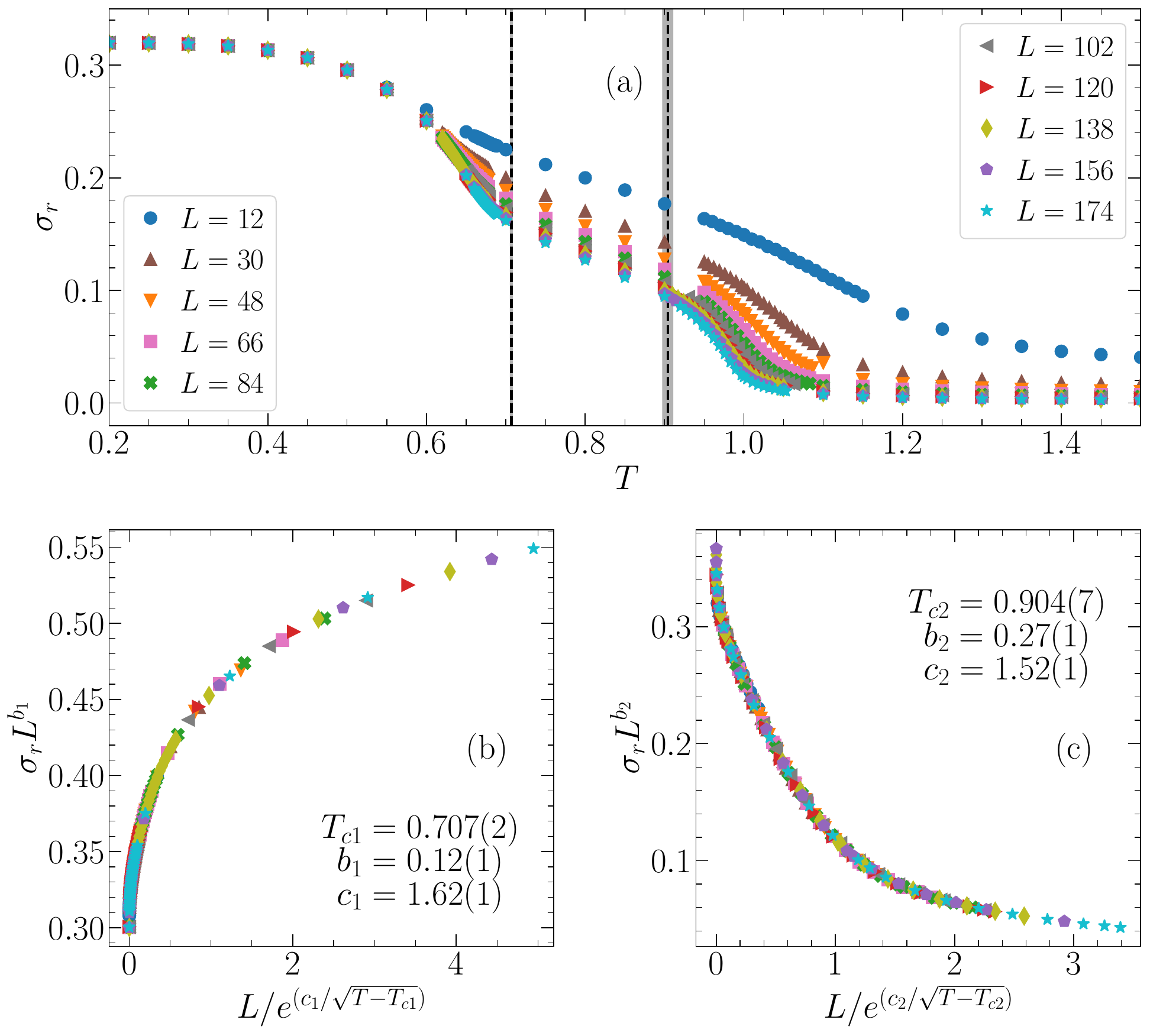}
\caption{(a) The standard deviation $\sigma_r$ with respect to temperature $T$ for the 6-clock state model on square lattices with different $L$. The vertical dashed lines mark the critical points from the finite-size scaling of $\sigma_r$, with the shaded area depicting the error. (b) and (c) show the best data collapse of the scaling for $T_{c1}$ and $T_{c2}$, respectively. }
\label{fig:sigma_q6}
\end{figure}

We continue to calculate the uncertainty $\sigma_r$ and perform the finite size scaling analysis with the BKT form following Eq.~\eqref{eq:scaling_BKT}. The temperature dependence of $\sigma_r$ for different sizes is illustrated in Fig.~\ref{fig:sigma_q6}(a). Guided by the different size dependencies in different regions, we carried out the scaling procedure separately close to the two possible transition points. Panels (b) and (c) in Fig.~\ref{fig:sigma_q6} show the data collapse around $T_{c1}$ and $T_{c2}$ respectively, and the obtained critical points agree well with previous studies~(see Table~\ref{tab:critical_information}).

\begin{figure}[!t]
\includegraphics[width=\columnwidth]{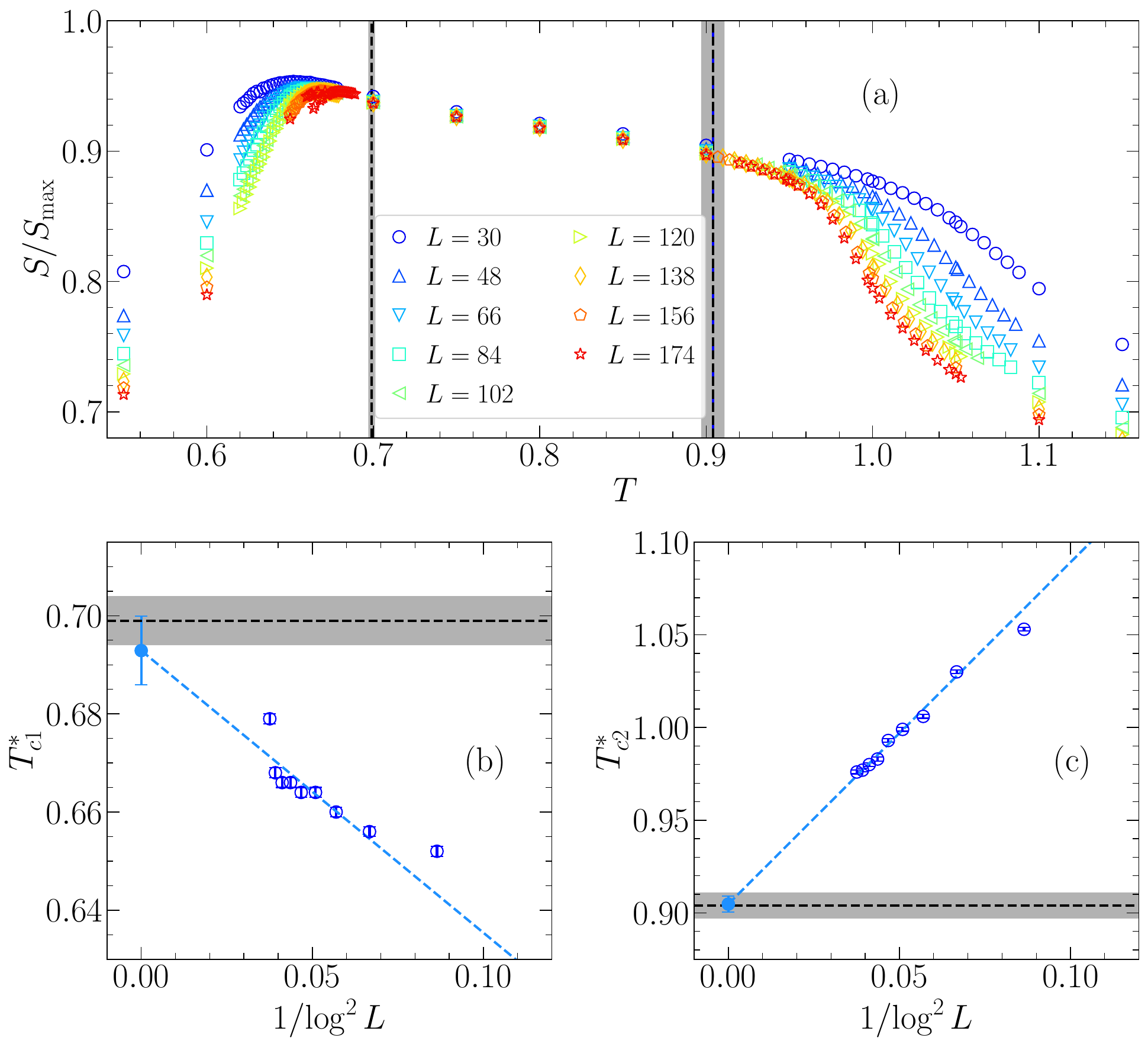}
\caption{(a) The participation entropy $S/S_\mathrm{max}$ verses $T$ for the 6-clock state model on square lattices with different $L$. The finite size extrapolation of $T_c^*$ is displayed in (b) for the first phase transition and (c) for the second, respectively. The vertical and horizontal dashed lines mark the critical point from the finite-size scaling in Fig.~\ref{fig:sigma_q6}, with the shaded area depicting the error.}
\label{fig:S_q6}
\end{figure}

In addition, we also check the temperature dependence of the participation entropy $S_r/S_\mathrm{max}$ for different sizes, as displayed in Fig.~\ref{fig:S_q6}(a). Here, $S_\mathrm{max}=Nq/2+1$ gives the upper bound of the participation entropy. The finite-size critical point $T_{c1}^*(L)$ can be extracted where $S$ reaches its maximum, and $T_{c1}^*(\infty)$ can be obtained through the finite-size extrapolation with the BKT scaling form. As displayed in Fig.~\ref{fig:S_q6}(b), considering the error bar, $T_{c1}^*$ in the thermodynamic limit from the extrapolation following the BKT form matches the first critical point got from the scaling of $\sigma_r$, which is denoted by the dashed line with shaded area for the error. The second finite-size critical point is extracted from the first-order deviation of $S$, and the extrapolated $T_{c2}^*$ also agrees well with the scaling results. Besides, in the critical phase between $T_{c1}$ and  $T_{c2}$, the participation entropy $S$ shows less system size dependence, and its value is larger than in the other two phases when $L\rightarrow \infty$.

\begin{figure}[!t]
\includegraphics[width=0.8\columnwidth]{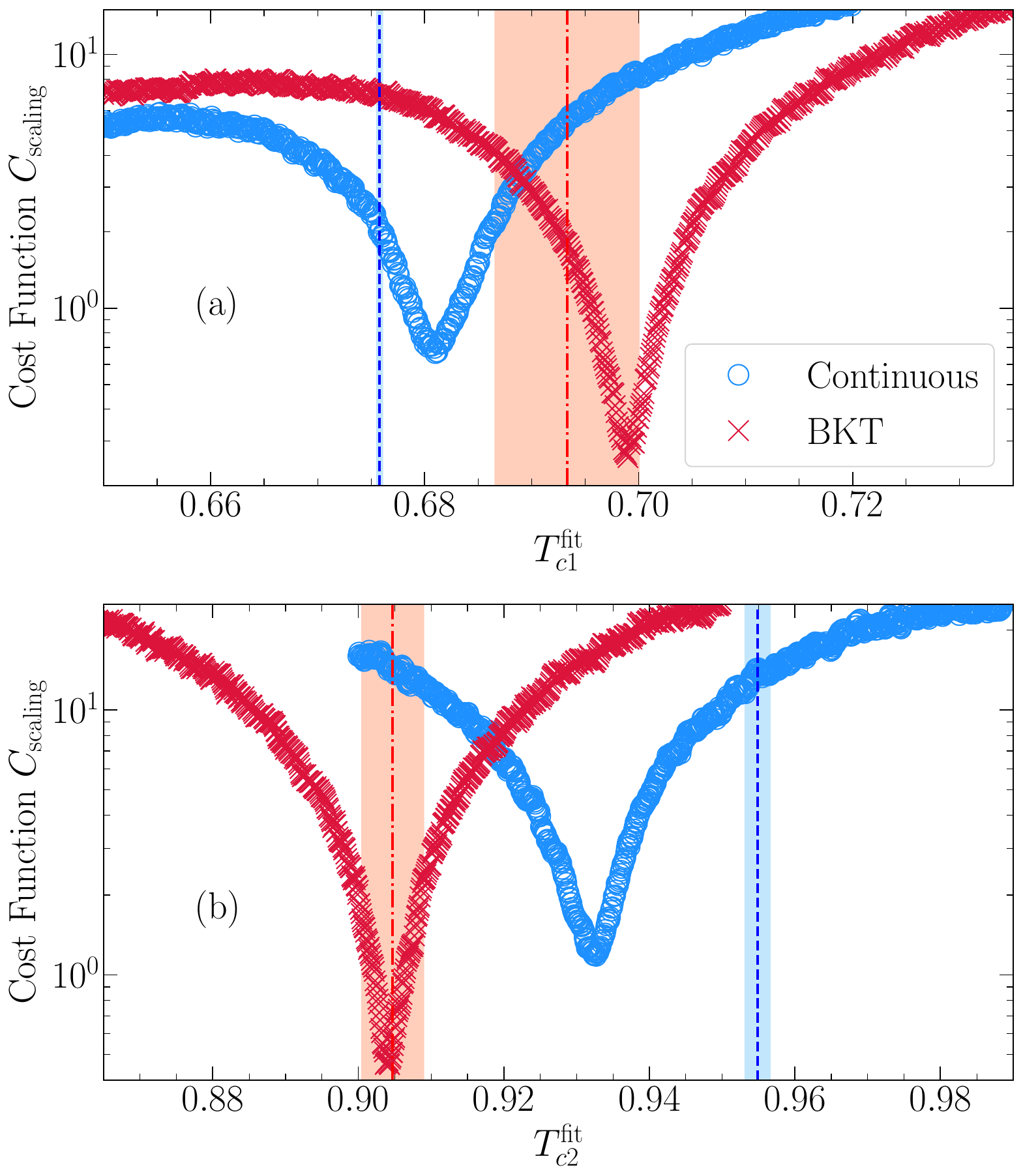}
\caption{The cost functions $C_\mathrm{scaling}$ in scaling procedures of $\sigma_r$ as a function of (a) $T_{c1}^\mathrm{fit}$ and (b) $T_{c1}^\mathrm{fit}$. The vertical dashed line depicts $T_c^*$ from the finite size extrapolation of $S_r$, with the shading area depicting the error. The blue and markers/lines depict procedures adopting the continuous and BKT type of phase transitions, respectively.}
\label{fig:transition_type_q6}
\end{figure}

While the finite size scaling analysis of $\sigma_r$ and $S$ in Fig.~\ref{fig:sigma_q6} and Fig.~\ref{fig:S_q6} strongly suggest the two BKT phase transitions, we also want to rule out the possibility of the continuous phase transition based on the analysis of $\{r\}$, assuming no prior knowledge about phases and phase transitions of the studied model. As displayed in Fig.~\ref{fig:transition_type_q6}, the cost function $C_\mathrm{scaling}$ is smaller in the scaling procedure with BKT form for both critical points, and the accordance between the scaling of $\sigma_r$ and extrapolation from $S_r$ is also better. In other words, the analysis based on the distances in the configuration space successfully determines the phase transition type without the help of prior information or other widely used physical quantities.   

\subsection{\texorpdfstring{$XY$}{Lg} model}

\begin{figure}[!b]
\includegraphics[width=\columnwidth]{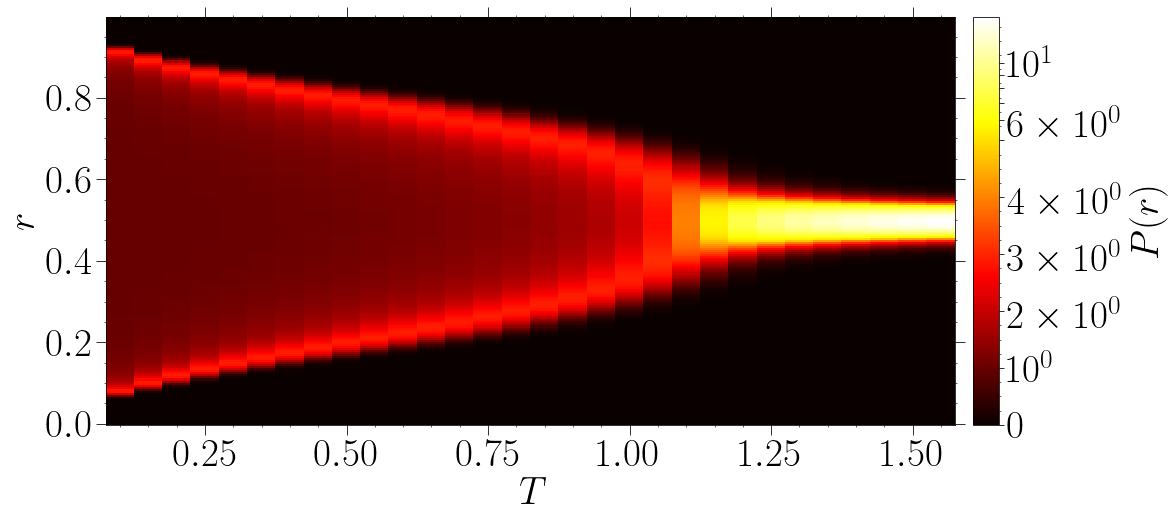}
\caption{ The distribution $P(r)$ as a function of temperature for the $XY$-model on a square lattice with the system length $L=174$.}
\label{fig:Pr_pcolor_xy}
\end{figure}

When $q\rightarrow\infty$, the $q$-state clock model becomes the $XY$ model, which is one of the first prototypical examples displaying topological long-range order and BKT phase transitions~\cite{Kosterlitz1973,Kosterlitz1974}. For the $XY$ model, the distance between the configurations defined by Eq.~\eqref{eq:distance_qstate} is continuous. It is reasonable to wonder whether the above procedure analyzing phase transitions remains functional in this case. 

We again perform the MC simulation on the $XY$ model and start from the data set of distances $\{r\}$ between the sampled configurations. Despite the continuous possible values, a numerical distribution can be obtained with discrete bins, which only affects the smoothness but not the value of the curve. As displayed in Fig.~\ref{fig:Pr_pcolor_xy}, the distribution $P(r)$ manifest two phases with one possible phase transition in the $XY$ model at finite temperature, and a further look shows that the low-temperature state of the $XY$ model exhibits continuous $P(r)$ similar to the intermediate phase in the $6$-state clock model. Besides, the ordered states with discrete symmetry shown in previous examples with several separated peaks in $P(r)$ disappear in the $XY$ model. In the high-temperature disordered phase, the $P(r)$ becomes a narrow distribution centered at $r=0.5$ too.

\begin{figure}[!t]
\includegraphics[width=\columnwidth]{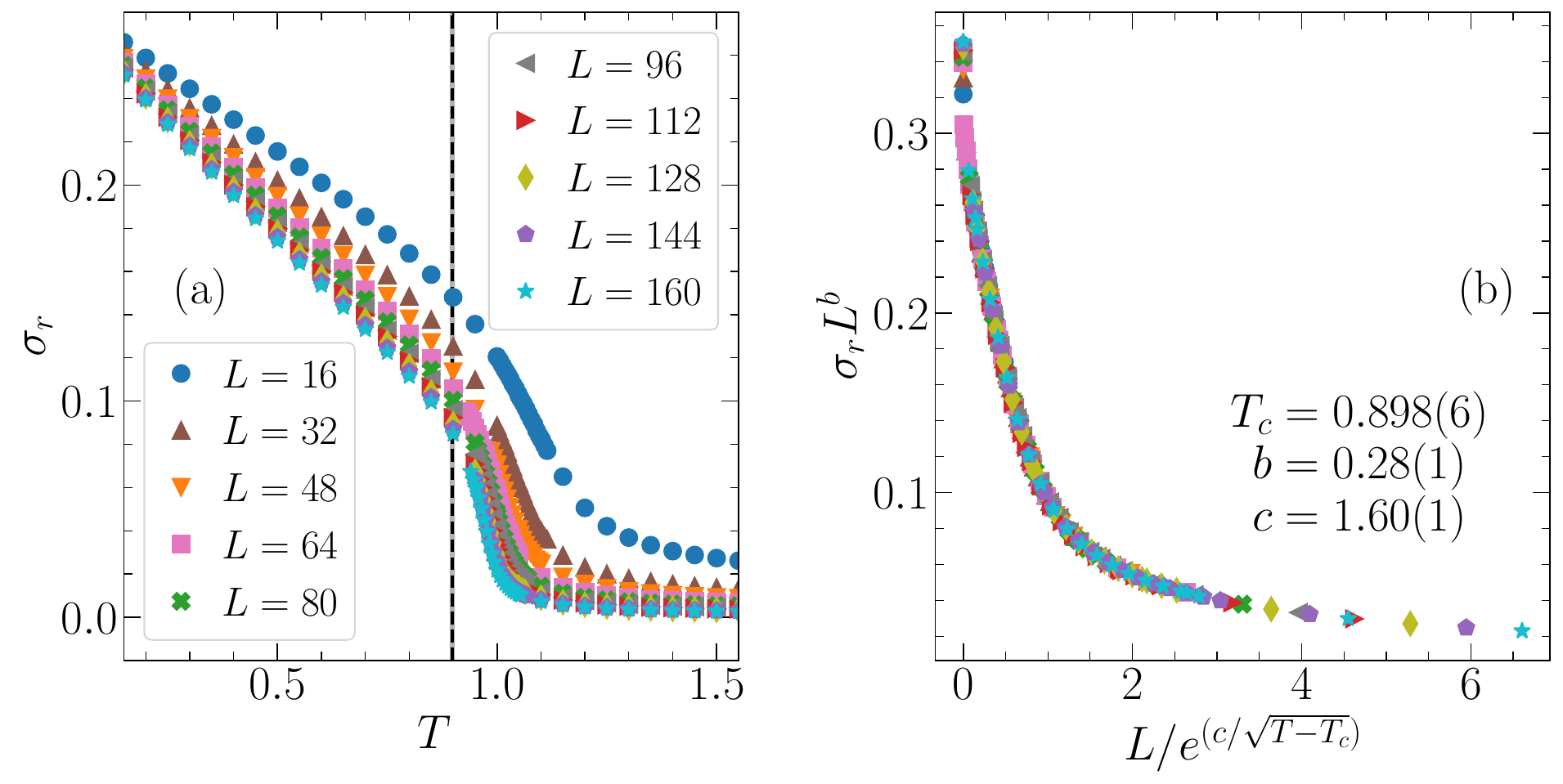}
\caption{ The standard deviation $\sigma_r$ with respect to temperature $T$ for the Ising model on square lattices with different $L$. The vertical dashed line marks the critical point from the finite-size scaling of $\sigma_r$, with the shaded area depicting the error.  (b) The best data collapse of the scaling with the obtained critical point and exponents.}
\label{fig:sigma_xy}
\end{figure}

The calculation of uncertainty is the same for discrete and continuous values of $r$, and the temperature dependence of $\sigma_r$ for different sizes is displayed in Fig.~\ref{fig:sigma_xy}(a). Adopting the scaling form in Eq.~\eqref{eq:scaling_BKT} for BKT phase transition, we can perform the data collapse of $\sigma_rL^{-b}$ versus $L/e^{c/\sqrt{T-T_{c}}}$ with all the different $L$, as for the $6$-state clock model. The best data collapse is shown in Fig.~\ref{fig:sigma_xy}(b), and the best scaling parameters therein agree with the representative literature (see Table~\ref{tab:critical_information}). 

\begin{figure}[!t]
\includegraphics[width=\columnwidth]{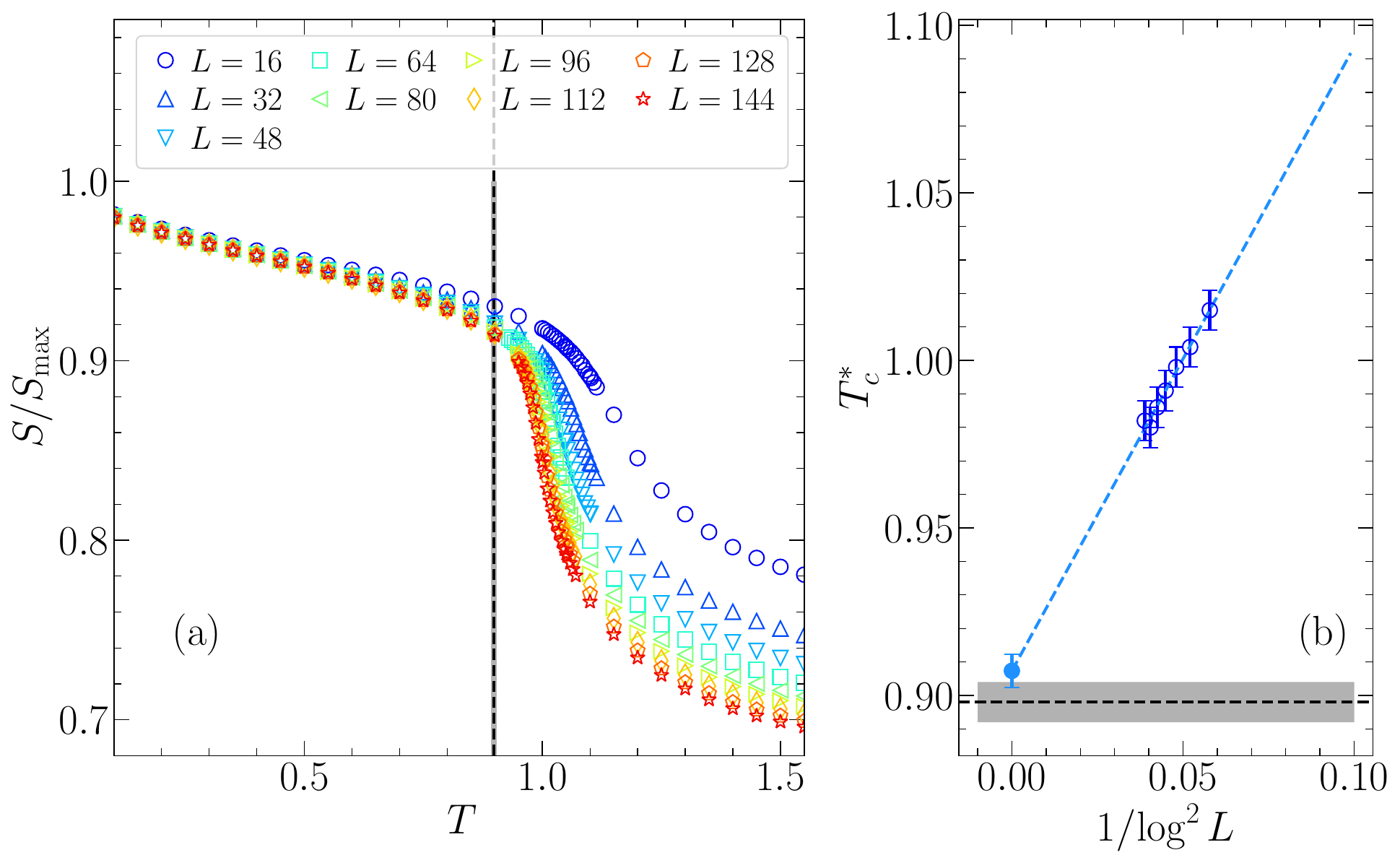}
\caption{(a) The participation entropy $S/S_\mathrm{max}$ verses $T$ for the $xy$ model on square lattices different $L$, with $S_\mathrm{max}=\log n_b$ and $n_b=100N$.  (b) The finite size extrapolation of $T_c^*$ points to the correct critical point at $1/L\rightarrow\infty$. The vertical/horizontal dashed line in (a)/(b) marks the critical point from the finite-size scaling in Fig.~\ref{fig:sigma_xy}, with the shaded area depicting the error.}
\label{fig:S_xy}
\end{figure}

The value of participation entropy can be sensitive to the way of discretization; namely, the value $S_r$ depends on the number of intervals $n_b$ when calculating $p(r)$ in the interval $[r-1/2n_b,r+1/2n_b]$. We briefly discuss the choice of $n_b$ in Appendix~\ref{appendix:xy_discretization} and choose $n_b=100N$ in Fig.~\ref{fig:S_xy}, which is sufficient for addressing the different behavior in different phases. We further extract the finite-size critical point from the first-order deviation of $S_r$ and perform the finite-size extrapolation with the scaling form of the BKT phase transition. As displayed in Fig.~\ref{fig:S_xy}(b), the extrapolated $T_c^*$ at the thermodynamic limit agrees well with the scaling results in Fig.~\ref{fig:sigma_xy}.   

We also want to emphasize that the temperature dependence of $\sigma_r$ and $S_r$ close to critical points show very similar behavior for the phase transition in the $XY$ model and the second critical point in the $6$-state clock model, even the two models have much difference in the symmetry of spins and the continuity of distances. While knowing that the two-phase transitions belong to the same universality class~\cite{youjin2022}, one would guess that the correlation in configuration space may further tell the universality class of the phase transition.

\section{critial exponent and the Universality class}
\label{sec:critical_exponent}

In all the systems studied in previous sections, the uncertainty $\sigma_r$ and the participation entropy $S_r$, extracted from the data set of the distances between the sampled configurations, successfully catch the phase transition points. We also obtain some critical exponents in the scaling procedure of $\sigma_r$. In this section, we further focus on the universal behavior of $\sigma_r$ and aim to figure out whether there is a connection between the uncertainty $\sigma_r$ and other known physical quantities. 

\begin{figure}[!t]
\includegraphics[width=\columnwidth]{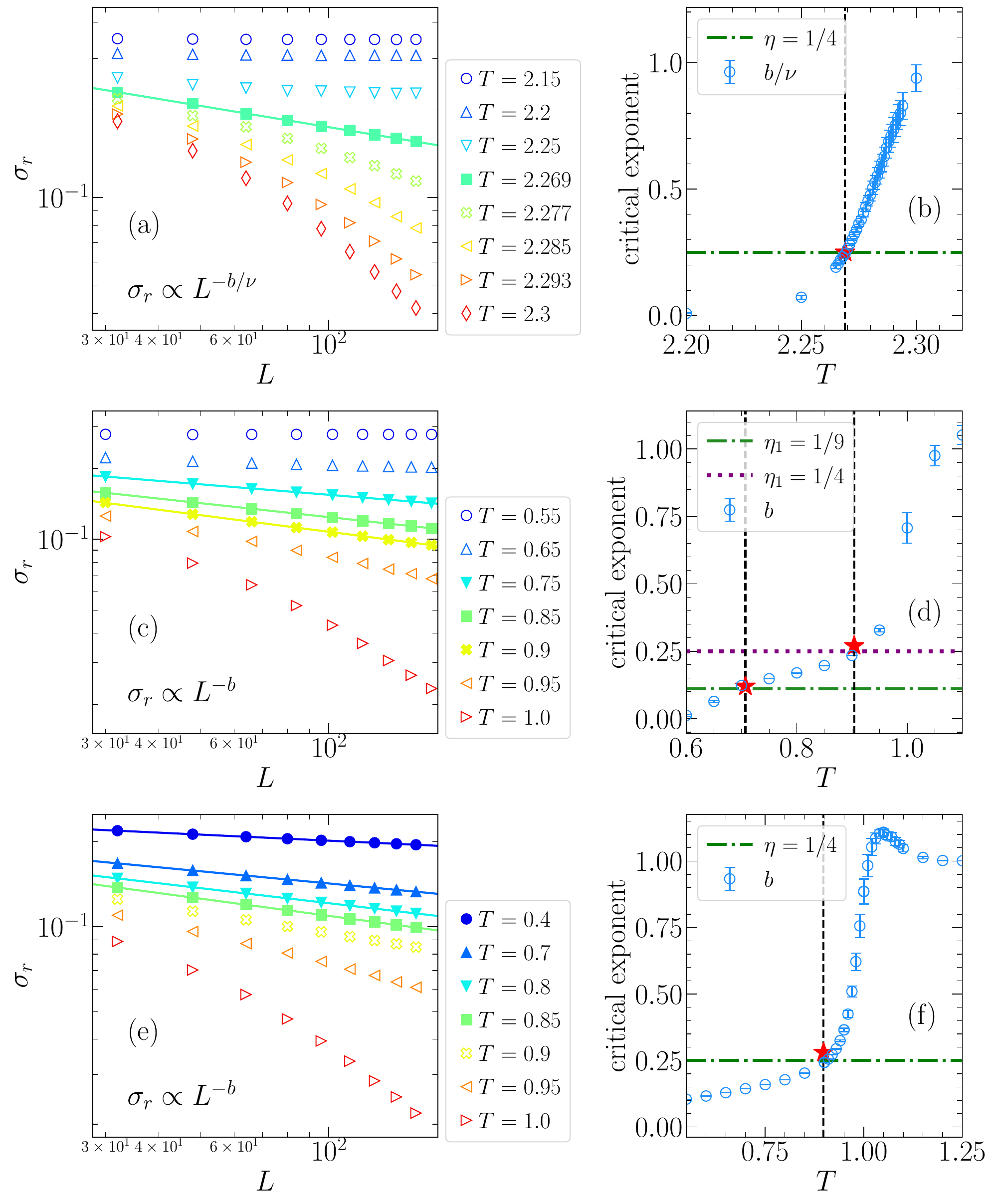}
\caption{The uncertainty $\sigma_r$ versus $L$ for (a) the Ising model, (c) $6$-statec clock model, and (e) $XY$ model, with the solid straight line depicting the power-law fitting. Right panels (b-e) show the exponent $b$ of power-law fitting $\sigma_r\propto L^b$. The red stars mark the critical points and exponents from the finite-size scaling of $\sigma_r$, the vertical-dashed lines depict the critical points from scaling, and the horizontal lines give the anomalous dimension. }
\label{fig:sigma_L}
\end{figure}

We begin our analysis with the size dependence of $\sigma_r$ close to the critical point, where the function ${\cal F}_\mathrm{con}(0)$ in Eq.~\eqref{eq:scaling_con} becomes a constant and one has $\sigma\propto L^{-b/\nu}$ for the continuous phase transition in the Ising model. As confirmed by the numerical results in Fig.~\ref{fig:sigma_L}(a), there is a clear power-law behavior of $\sigma_r$ versus $L$. In the ordered phase at low temperatures, $\sigma_r$ shows minor size dependence, while for temperatures slightly larger than $T_c$, the uncertainty decays much faster than the power law. Nevertheless, we perform a power-law fitting of $\sigma_r$ versus $L$ for all temperatures and plot the power-law exponent in Fig.~\ref{fig:sigma_L}(b). First, the obtained exponent $b/\nu$ agrees well with the $[T_c,b/\nu]$ from the finite-size scaling, which is marked by the red star. Comparing our results with the known critical exponent of the Ising model, we find that $b/\nu$ numerically equals the anomalous dimension $\eta$, which determines the real-space decay of correlation functions and is depicted by the horizontal dash-dotted line. 

For the BKT phase transition, one obtains $\sigma_r\propto L^{-b}$ close to the critical point with a similar analysis. In Fig.~\ref{fig:sigma_L}(c) and (e), we display the $L$-dependence of $\sigma_r$ in the double-logarithmic scale, for the $6$-state clock model and the $XY$-model, respectively. In both cases, the uncertainty $\sigma_r$ shows power-law decay behavior in the quasi-long-range-ordered critical phase, and the critical exponent obtained from the power-law fitting at each critical point matches the scaling results. Moreover, for each transition point, the value of exponent $b$ is again very close to the known anomalous dimension $\eta$. While the same $\eta=1/4$ for the phase transition in the $XY$ model and second critical point in the $6$-state clock model demonstrate the same universality class, the critical exponent $b$ related to $\sigma_r$ is very likely to have the same scaling behavior. 

\begin{figure}[!t]
\includegraphics[width=\columnwidth]{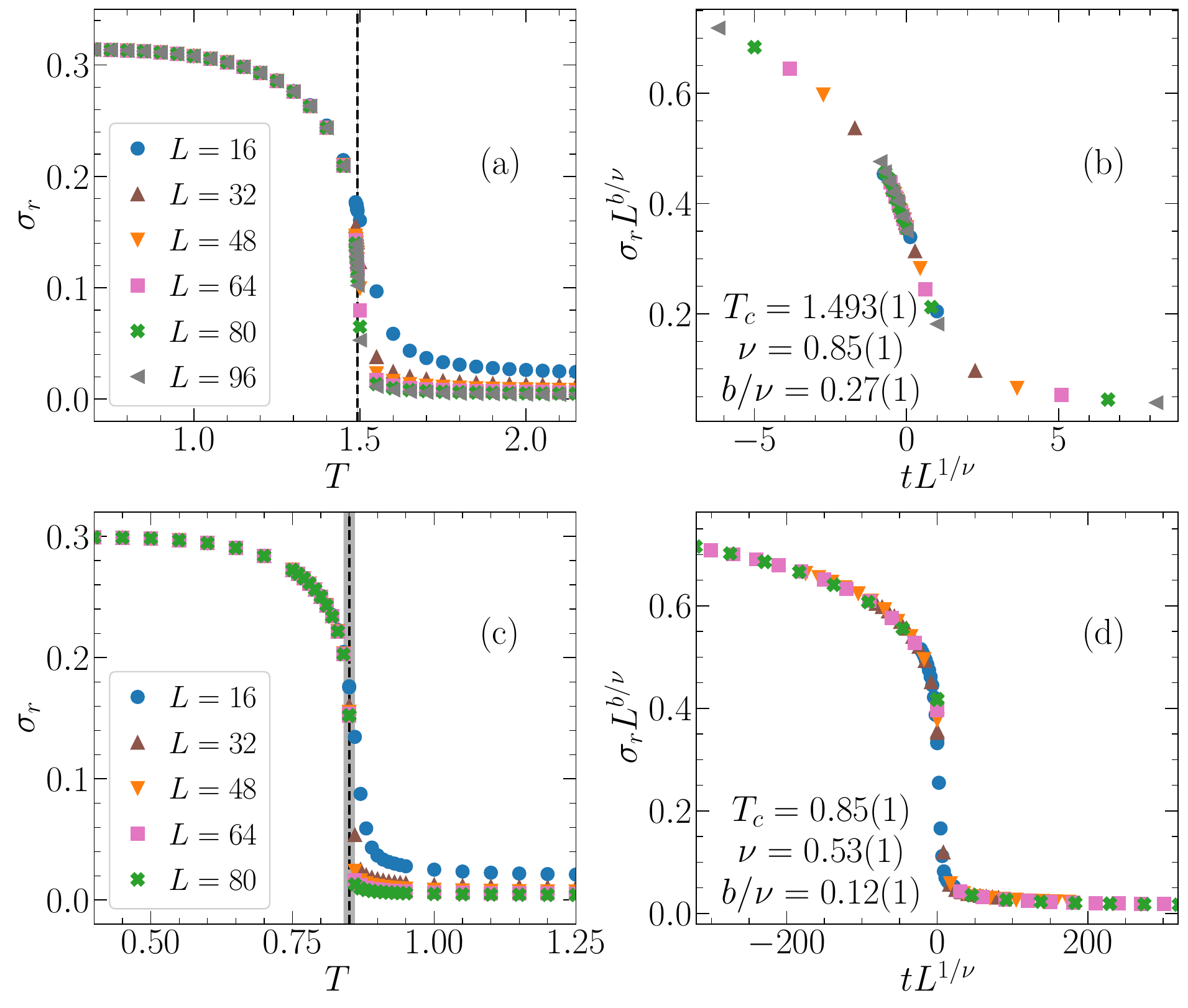}
\caption{ The standard deviation $\sigma_r$ with respect to temperature $T$ for (a) the $3$-state clock model and (c) the $5$-state Potts model. The vertical dashed line marks the critical point from the finite-size scaling of $\sigma_r$, with the shaded area depicting the error. The right panels show the best data collapse of the scaling with the obtained critical point and exponents for (c) the $3$-state clock model and (d) the $5$-state Potts model. }
\label{fig:sigma_others}
\end{figure}

To verify our conjecture that the critical exponent of $\sigma_r$ equals the anomalous dimension $\eta$, we further extend our analysis to the $3$-state clock model and the $5$-state Potts model. The $q$-Potts model shares the same local sites with the $q$-state, and is described by the Hamiltonian
\begin{align}
    H_\mathrm{Potts} = -J\sum_{\langle ij\rangle} \delta_{\theta_i \theta_j}. 
\end{align}
While the $3$-state clock model features a continuous phase transition and the $5$-state Potts model features a first-order phase transition, one can anyway perform the same finite-size analysis as for the continuous phase transition, and the obtained critical exponent would tell the difference. The uncertainty $\sigma_r$ and the corresponding finite-size scaling are displayed in Fig.~\ref{fig:sigma_others}, and the data collapse is of good quality. In Fig.~\ref{fig:sigma_others}(d), the correlation length exponent $\nu=0.53(1)$ is very close to $1/2$ for the $5$-state Potts model, indicating a first-order phase transition. 

\begin{figure}[!t]
\includegraphics[width=\columnwidth]{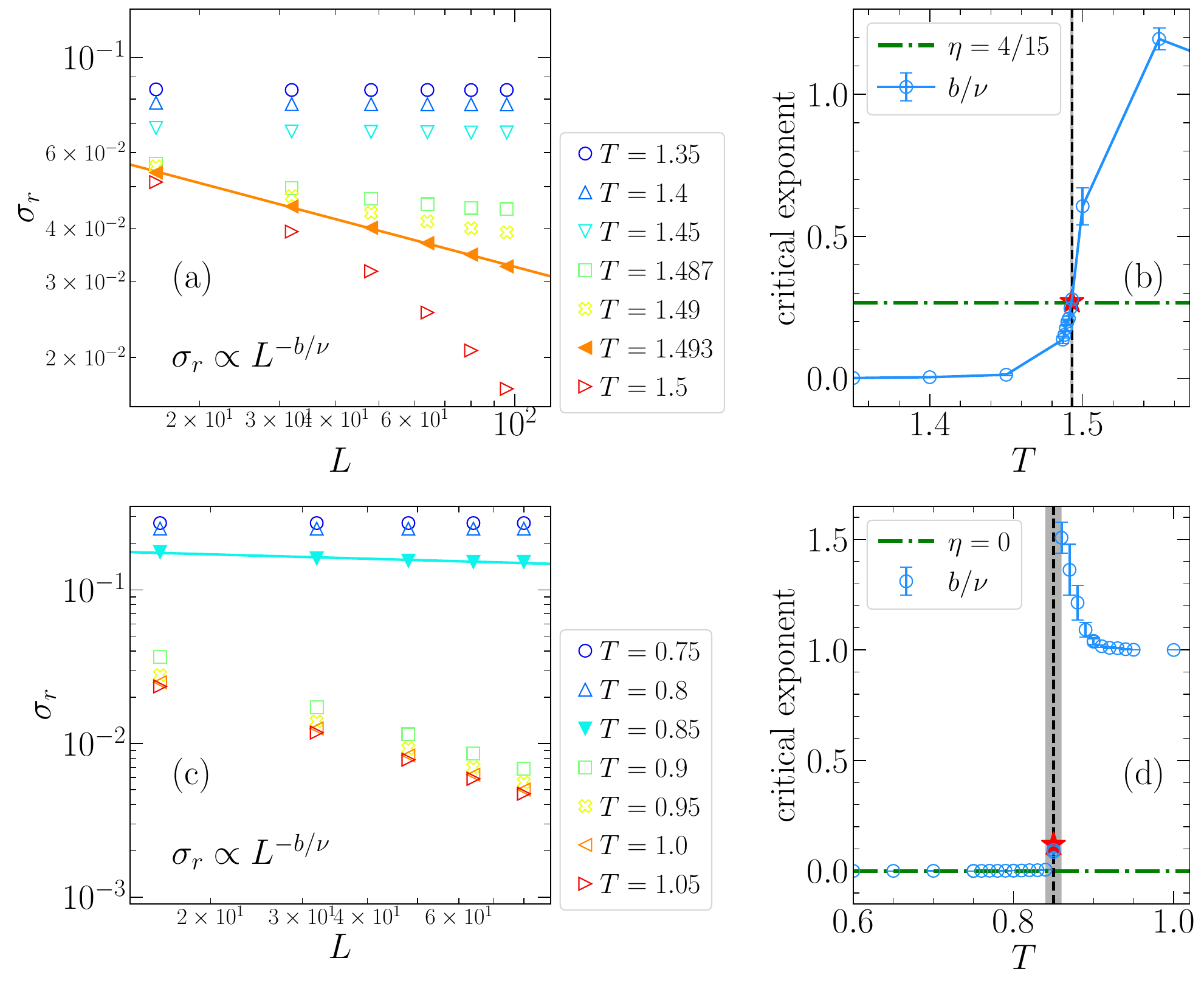}
\caption{The uncertainty $\sigma_r$ versus $L$ for (a) the $3$-state clock model, and (c) $5$-state Potts model. Right panels (b-d) show the exponent $b$ of power-law fitting $\sigma_r\propto L^b$. The red stars mark the critical points and exponents from the finite-size scaling of $\sigma_r$, the vertical-dashed lines depict the critical points from scaling, and the horizontal lines give the anomalous dimension. }
\label{fig:sigma_L_others}
\end{figure}

For the $3$-state clock model, the power-law decay of $\sigma_r$ versus $L$ is apparent at the critical point of a continuous phase transition, as shown in Fig.~\ref{fig:sigma_L_others}(a). Figure~\ref{fig:sigma_L_others}(b) demonstrate the great consistency among $b/\nu$ from the power-law fitting in Fig.~\ref{fig:sigma_L_others}(a), $b/\nu$ from the scaling in Fig.~\ref{fig:sigma_others}(b), and the known anomalous dimension $\eta=4/15$. For the $5$-state Potts model, $\sigma_r$ versus $L$ in Fig.~\ref{fig:sigma_L_others}(c) shows distinguish behavior at low and high temperatures. The critical exponent extracted from the power-law fitting changes dramatically near the critical points, which results in a considerable mismatch between $b/\nu=0.12(1)$ with the anomalous dimension $\eta=0$. This mismatch can be eliminated by considering an infinitesimal shift of $T_c$ to the low-temperature ordered phase, and one has robust $b/\nu\rightarrow 0$.

\begin{figure}[!b]
\includegraphics[width=\columnwidth]{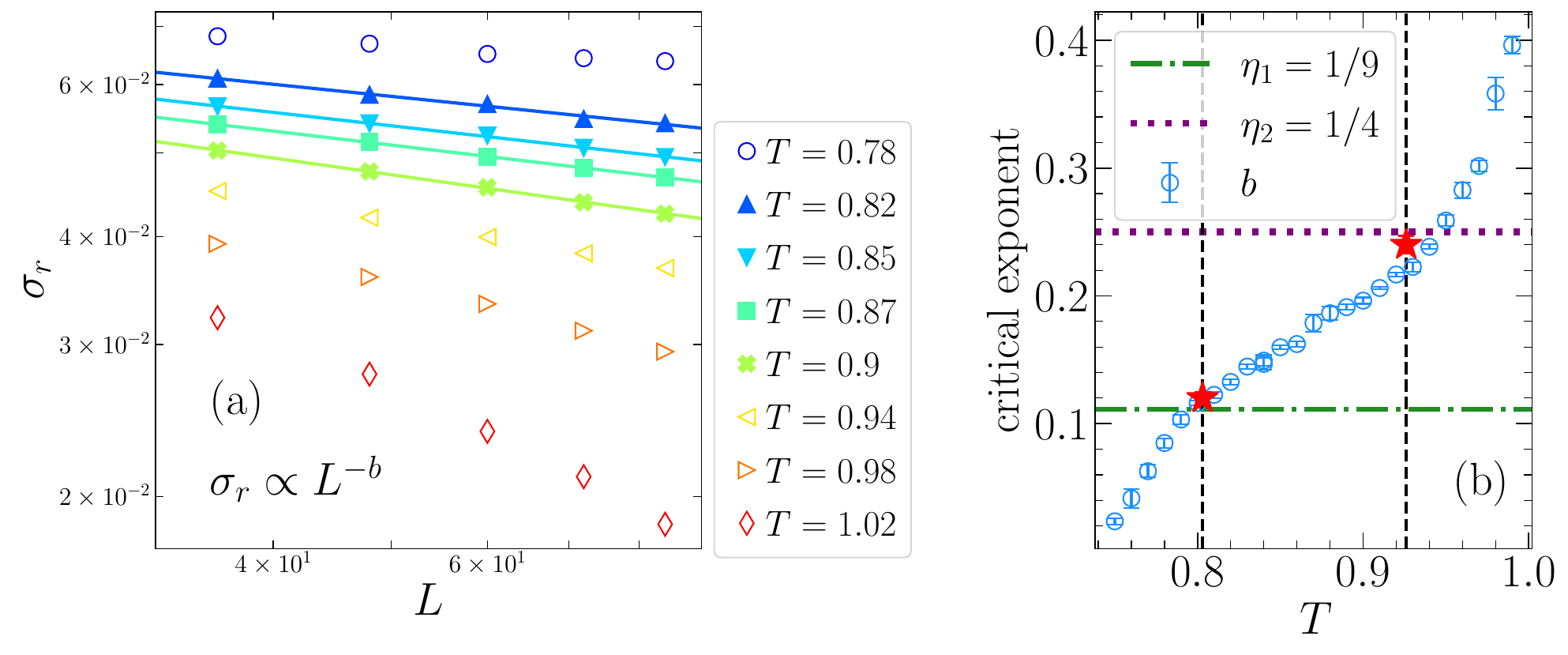}
\caption{(a) The uncertainty $\sigma_r$ versus $L$ for KIM. (b) The critical exponent $b$ of power-law fitting $\sigma_r\propto L^b$. The red stars mark the critical points and exponents from the finite-size scaling of $\sigma_r$, the vertical-dashed lines depict the critical points from scaling, and the horizontal lines give the anomalous dimension. }
\label{fig:sigma_L_kim}
\end{figure}

While the above systems have different local degrees of freedom and different phase transitions, they live on the same square lattice with the same order parameter, i.e., magnetization. To verify the generality of our analysis, we further check the kagome Ising model (KIM) with the complex order parameter (See appendix \ref{appendix:kim}). With properly chosen parameters, KIM supports the six-fold spin ice ground state and has two consecutive BKT phase transitions, which belong to the same universality class as the 6-state clock model~\cite{Su2023,chern2012, Wolf_kagome_1988}. As displayed in Fig.~\ref{fig:sigma_L_kim}, the system size dependence of $\sigma_r$ is quite similar to the case of $6$-state clock model, and the extracted critical exponents at two critical points match the predicted anomalous dimension. In other words, although the local sites, the lattices, and the order parameters are different, the same analysis based on the distances between the sampled configurations catches the same universal information of the phase transition. 

\begin{table}[!t]
    \centering
    \begin{tabular}{c  c  c}
        \hline
        \hline
                & Ising model  &           \\
        \hline 
        present work  & $T_c=2.269(1)$   & $b/\nu=0.25(1)$ \\
        \hline 
      exact~\cite{Sandvik2010} & $T_c\approx 2.269$ & $\eta=1/4$      \\
        \hline
        \hline
                & 6-state clock model &           \\
        \hline 
        present work  &  $T_{c1} = 0.707(2) $  &  $b_1 = 0.12(1)$ \\
                      &  $T_{c2} = 0.904(7) $  &  $b_2 = 0.27(1)$ \\
        \hline 
         exact~\cite{rnsix1}&  &  $\eta_1 = 1/9$    \\
                                      &  & $\eta_2=1/4$      \\
         MC~\cite{crisix1}& $T_{c1} = 0.7014(11)$ &  $\eta_1 = 0.113(1)$     \\
                                      & $T_{c2} = 0.9008(6)$ & $\eta_2=0.243(4)$     \\                             
      TNS~\cite{Li2022_qstateclock} & $T_{c1} = 0.696(2)$ &       \\
                                      & $T_{c2} = 0.9111(5)$ &       \\
        \hline
        \hline
                & $XY$ model  &           \\
        \hline 
        present work  & $T_c=0.898(6)$   & $b=0.28(1)$ \\
        \hline 
         exact~\cite{Kosterlitz1973} &  & $\eta=1/4$      \\
          MC~\cite{Hsieh_2013} & $T_{c}=0.8935(1)$ &  \\
     TNS~\cite{tcforxy} & $T_{c}= 0.89290(5)$  &      \\
        \hline
        \hline
                & $3$-state clock model  &           \\
        \hline 
        present work  & $T_c=1.493(1)$   & $b/\nu=0.27(1)$ \\
        \hline 
        exact~\cite{potts3,zhenggu} & $T_{c}\approx1.49246$ & $\eta=4/15$\\
        MC~\cite{Pascal}&& $\eta=0.2755(6)$\\
        TNS~\cite{zhenggu}& $T_{c}\approx1.4925(3)$ & \\

        \hline 
        \hline
                & $5$-state Potts model  &           \\
        \hline 
        present work  & $T_c=0.85(1)$   & $b=0.12(1)$ \\
        \hline 
      exact~\cite{q=5_Potts} & $T_{c}\approx 0.85153$& $\eta=0$     \\
      MC~\cite{q=5_Potts}& $T_{c}=0.851(5)$ &  $\eta=0.230(8)$\\

        \hline
        \hline
                        & Kagome Ising model  &           \\
        \hline 
        present work  & $T_{c1}=0.809(2)$   & $b_1=0.12(1)$ \\
                      & $T_{c2}=0.927(7)$   & $b_2=0.24(1)$ \\
        \hline 
      exact~\cite{Wolf_kagome_1988} &  & $\eta_1=1/9$     \\
                                    &  & $\eta_2=1/4$     \\
      MC~\cite{Su2023}& $T_{c1}=0.803(5)$ &  \\
      & $T_{c2}=0.926(7)$ &  \\
        \hline
        \hline
    \end{tabular}
    \caption{A summary of critical points and exponents for the models studied in the present work, compared with the representative literature.}
    \label{tab:critical_information}
\end{table}

We summarize the useful critical information from the finite-size scaling of $\sigma_r$ in Tab.~\ref{tab:critical_information}, with a comparison to results in some representative literature. First, our scaling procedure produces quite good critical points in all systems with different kinds of phase transitions. Second, based on our numerical results, we can make the following conjecture: the critical exponent of $\sigma_r$, i.e., $b/\nu$ for the continuous phase transition or $b$ for the BKT phase transition, equals the anomalous dimension $\eta$. While the anomalous dimension is related to the decay of the correlation function in real space, $\sigma_r$ is a function of distances between sampled configurations and can be considered as a generalized correlation in the configuration space. Whether the correlations in real space and Hilbert space have deeper connections, and how to explain the equality between $b$ (or $b/\nu$) and $\eta$, are beyond the present work and deserve more investigations. 

\section{Summary and discussion}
\label{sec:sum}

Based on the sampled configurations and the 1-norm distances between them, we propose a new way of studying phase transitions in the lattice models. The very basic data analysis of the distances $\{r\}$, such as the probability $p(r)$, the uncertainty $\sigma_r$, and the Shannon entropy $S_r$, can already provide us with rich information on the phase transitions. Specifically, using the sampled configurations from the numerical MC simulations, $\sigma_r$ and $S_r$ are demonstrated to follow the finite-size scaling behavior. The corresponding finite-size analysis catches the phase transition point and the phase transition type in a series of statistical lattice models, including the Ising model, $q$-state clock models, $q$-state Potts model, the $XY$ model on the square lattice, and KIM. Moreover, in all these numerical experiments, the critical exponent $b$ (or $b/\nu$ for the continuous phase transition) related to the uncertainty $\sigma_r$ is found to be very close to the anomalous dimension $\eta$, which determines the scaling behavior of the correlation function in the real space. 

The numerical results involve different systems with different numbers of local degrees of freedom, different numbers of phases, different phase transition types, and different lattice geometries. In all cases, the simple analysis based on the distances between sample configurations shows a universal critical behavior, and the critical information is extracted unbiasedly without the help of the order parameter or other widely used physical quantities. Compared to the recent ML-related works~\cite{santos1, Santos2021prxq, Sale2022}, which share a similar idea of investigating phase transition with the MC sampled configurations, our proposal is much more straightforward and transparent. On the other hand, while the ML language prefers the data-mining process as black boxes, we consider the function of distances as the generic correlation in the Hilbert space, which should have its own physical significance and deserve further investigation.  

The present does not aim for more accurate critical points or exponents with more digits of the studied models. Our main goal is to demonstrate that the sampled configurations are sufficient to probe phase transitions, and we propose the simplest way to do it with correlations in Hilbert space. Since the procedure is system-independent and requires no foreknowledge of the phase or phase transition, we hope our proposal can shed light on the complicated phase transitions in complex systems without a proper order parameter, such as the unknown phase transitions in the HoAgGe compound~\cite{KZhao}, which can be described by the distorted kagome Ising model with long-range interactions. Moreover, while the MC simulation is the most widely used sampling method, the quantum collapse measuring in the rapidly developed quantum computing experiment provides an alternative way of sampling for closed quantum systems. Since the Hilbert space and the distance are the same as in the quantum lattice models, our proposal has the potential to be implemented in studying quantum phase transitions, especially in quantum computing experiments. 

\section*{Acknowledgments}
This research was supported by the National Natural Science Foundation of China (grant nos. 12174167, 12004020, 12247101), and the Fundamental Research Funds for the Central Universities.

\appendix

\section{Distances in the high-temperature limit}
\label{appendix:Pr_analysis}

At the infinite temperature, the configurations are uniformly distributed, and the probability of distances is explicitly determined by Eq.~\eqref{eq:Pr_inf_T}. In this case, $r_{\alpha,\beta}$ in Eq.~\eqref{eq:distance_qstate} between an arbitrary pair is the average of $N$ independent and identically distributed variables $r_i=r(\theta_i^\alpha,\theta_i^\beta)/\pi$. According to the central limit theorem, as long as the number of variables is large enough, the average of independent and identically distributed random variables can be approximated to a normal distribution with average $\overline{r}=\overline{r_i}$ and the variance $\sigma^2_r=\sigma^2_{r_i}/N$. Therefore, for the lattice models with two on-site levels, such as the Ising model, distance $R$ in Eq.~\eqref{eq:distance_q2} follows a binomial distribution $B(N/2,N/4)$. When $N$ is large enough, $R$ approximately follows a normal distribution with the mean value $N/2$ and the variance $N/4$. Then the normalizad distance $r=R/N$ also follows a normal distribution, with $\overline{r} = 1/2$ and
\begin{align}
    \sigma^2_r = \frac{1}{4N}.
\end{align}

For the participation entropy in Eq.~\ref{eq:entropy}, it should be noted that some literature has used differential entropy as the near continuous limit for participation entropy. Unlike the participation entropy, the differential entropy is not in general a good measure of uncertainty or information. For example, the differential entropy can be negative; it is also not invariant under continuous coordinate transformations. In order to more accurately estimate the value of participation entropy, we use limiting density of distinct points (LDDP)~\cite{JayneLDDP1968} instead of differential entropy as
\begin{align}
    S_r &=- \sum_{r} p(r) \ln(p(r)) \notag \\
    &\approx -\int_{-\infty}^{\infty} P(r) \ln(\frac{P(r)}{m(r)}) dr. 
\end{align}
Here, $m(r)$ is the distribution density function of discrete points, defined as
\begin{equation}
    \lim_{N_r\to\infty} \{\mathrm{No.}~r~\mathrm{in}~(a,b)\}=\int_a^b m(r)dr.
\end{equation}   
Notice that $r$ is normalized to $[0,1]$ with $N_r$ equally spaced values, one has $m(r) = 1/N_r$. Therefore, we can estimate the participation entropy as
\begin{equation}
    S_r \approx \ln(N_r) +\frac{1}{2} \ln(2\pi e\sigma_r^2).
\end{equation}
Since the lattice models two-level sites has $N_r=N+1$, its participation entropy reads
\begin{align}
    S_r\approx \frac{1}{2}\ln(N)+\frac{1}{2}\ln(\pi e/2).
\end{align}

For $q$-state clock models with $q \geq 4$, there are $N_r =  q/2+1$ possible values for $r_i$, and its probability reads
\begin{align}
    p(r_i)=
    \begin{cases}
    1/q& r_i=0~\mathrm{or}~1,\\ 
    2/q& \mathrm{else},
    \end{cases}
\end{align}
from which one easily obtains $\overline{r_i} = \sum p(r_i) r_{i} = 1/2$, $\overline{r_i^2} = \sum p(r_i) r^2_{i} = 1/3 + 2/3q^2$, and $\sigma^2_{r_i} =\overline{r_i^2}-(\overline{ r_i})^2=1/12+2/3q^2$. And for the distance between configurations, we have $\overline{r} = 1/2$,
\begin{align}
    \sigma^2_r =& \frac{1}{12N}+\frac{2}{3Nq^2}, \\
    S_r \approx& \frac{1}{2}\ln(N)+\frac{1}{2}\ln(\pi e(q^2+8)/24). \nonumber
\end{align}

In the limit $q\to\infty$, $\theta^\alpha_i$ on each site approximately follow a uniform distribution on $[0,2\pi]$, and  $r_i$ follows a uniform distribution $ U(0,1)$. One therefore has $\overline{r} = 1/2$ and
\begin{align}
    \sigma^2_r =& \frac{1}{12N}.
\end{align}   
The participation entropy for the continuous $r$ in the $XY$ model will be discussed in detail in Appendix~\ref{appendix:xy_discretization}.

The analysis in the section depends only on the nature of the local site but not the details of the system. For the square lattice in the main text, one easily obtains the $L$ dependence of the $r$-related quantities by adopting $N=L^2$.

\section{Simulation details}
\label{appedex:sampling_details}

The MC simulation in the present work combines the Metropolis algorithm~\cite{Metropoliesupdate}, Wolff algorithm~\cite{Wolffupdata}, and the annealing process~\cite{Annealingprocess}. In each MC step, we perform a Metropolis update and a Wolff update to quickly reach the equilibrium for both low and high temperatures. In addition, the simulation of a lower temperature is guided by the results of a higher one, i.e., the annealing process is adopted for better ergodicity and convergence. 

For all calculations with different system sizes, we take $N_b = 56$ independent bins of MC procedures; Each bin has 10000 MC steps to reach equilibrium and 20000 MC steps for measurement. In the measurement part, we take one configuration per 100 steps and $N_c=200$ configurations in each bin. These settings are sufficient to avoid the autocorrelation between the two MC steps and make ergodic samplings. In total, We have a data set $\{s\}$ with ${\cal N}_s=N_c\times 56\approx 10^5$ configurations. Then the data set $\{r\}$ is obtained by computing distances between all possible pairs of configurations, which results in a total number of distances as ${\cal N}_r = {\cal N}_s({\cal N}_s-1)/2 \approx 10^{10}$. With the data set $\{r\}$, one easily obtains $P(r)$ from a simple histogram, the average and the uncertainty according to Eq.~\eqref{eq:r_mean_sigma}, and the Shannon entropy according to Eq.~\eqref{eq:entropy}. The computational costs related to $\{s\}$ and $\{r\}$ are negligible in the whole MC simulation process. 

In the above procedure, we collect all configurations from different bins to generate target measurements, i.e., $P(r)$, $\sigma_r$ and $S_r$. The mix of different bins is very important for ergodicity since the degenerate samples of configuration can produce misleading measurements in a single bin. For example, at sufficiently low temperatures, a single measuring of the Ising model collapses to one of the two degenerated ground states, which leads to a delta peak of $P(r)$ randomly at $r=0$ or $1$ with $\sigma_r=0$ and $S_r=0$. Further averaging these measurements of different bins is meaningless, and the results can not catch the key feature of correlations in configuration space, as demonstrated in Fig.~\ref{fig:schematic}(b). 

Unlike traditional observables from typical MC simulations, where one gets a statistical error from parallel bins, one would notice that this procedure generates no statistical error with mixed bins. To include the statistical error from different data collections, one can increase the number of bins $N_b=N_{b0}\times N_{b1}$, and mix every $N_{b0}$ bins to have meaningful measurements in configuration space, and further averaging $N_{b1}$ measurements with statistical errors. Alternatively, one can also divide each bin into many sequential sections, and sections in different bins with the same sequence index can form an ergodic data collection. Nevertheless, we do not have statistical errors in the present work, which does not affect our results and story. 

\section{Scaling procedure and Data collapse}
\label{appedix:scaling_details}

As mentioned in the main text, the critical $T_c$, $b$, and $\nu$ in the scaling form of Eq.~\eqref{eq:scaling_con} is numerically determined by the best data collapse, where the quality of data collapse is estimated by the cost function defined as~\cite{Jan2020,Aramthottil2021,liang2023disorder,Su2023} 
\begin{align}
    C_{\mathrm{scaling}}=\frac{\sum_j |Y_{j+1}-Y_j|}{\max\{Y_j\}-\min\{Y_j\}}-1.
    \label{eq:cost_function}
\end{align}
Here we use $[X,Y]$ storing the data in Fig.~\ref{fig:sigma_q2}(b), with $X$ and $Y$ correspond to $tL^{1/\nu}$ and $\sigma_r L^{-b/\nu}$. By sorting $X_j$ in a nondecreasing way with $X_j\leq X_{j+1}$, the minimum $C_{\mathrm{scaling}}$ gives the smoothest curve. 

In the present work, we assume no foreknowledge of the critical parameters and perform brute force search in the three-dimensional parameter space $\{T_c, b, \nu \}$. In other words, we compute $C_{\mathrm{scaling}}$ on a three-dimensional grid of $\{T_c^{\mathrm{fit}}, b^{\mathrm{fit}}, \nu^{\mathrm{fit}} \}$, the the critical parameters are determined at where the minimum $C_{\mathrm{scaling}}$ appears. 

As mentioned in Sec.~\ref{appedex:sampling_details}, the input data $\sigma_r$ has no statistical error, and the definition in Eq.~\eqref{eq:cost_function} contains no error as well. Therefore, we estimate the error of the critical parameters as their minimum intervals on the grid; for the critical temperature, we consider both the interval of the scaling grid and the simulation temperature, and the larger of them is set as the error. The scaling with the BKT form in Eq.~\eqref{eq:scaling_BKT} follows the same procedure. 

\section{Discretization in the \texorpdfstring{$XY$}{Lg} model}
\label{appendix:xy_discretization}

\begin{figure}[!t]
\includegraphics[width=\columnwidth]{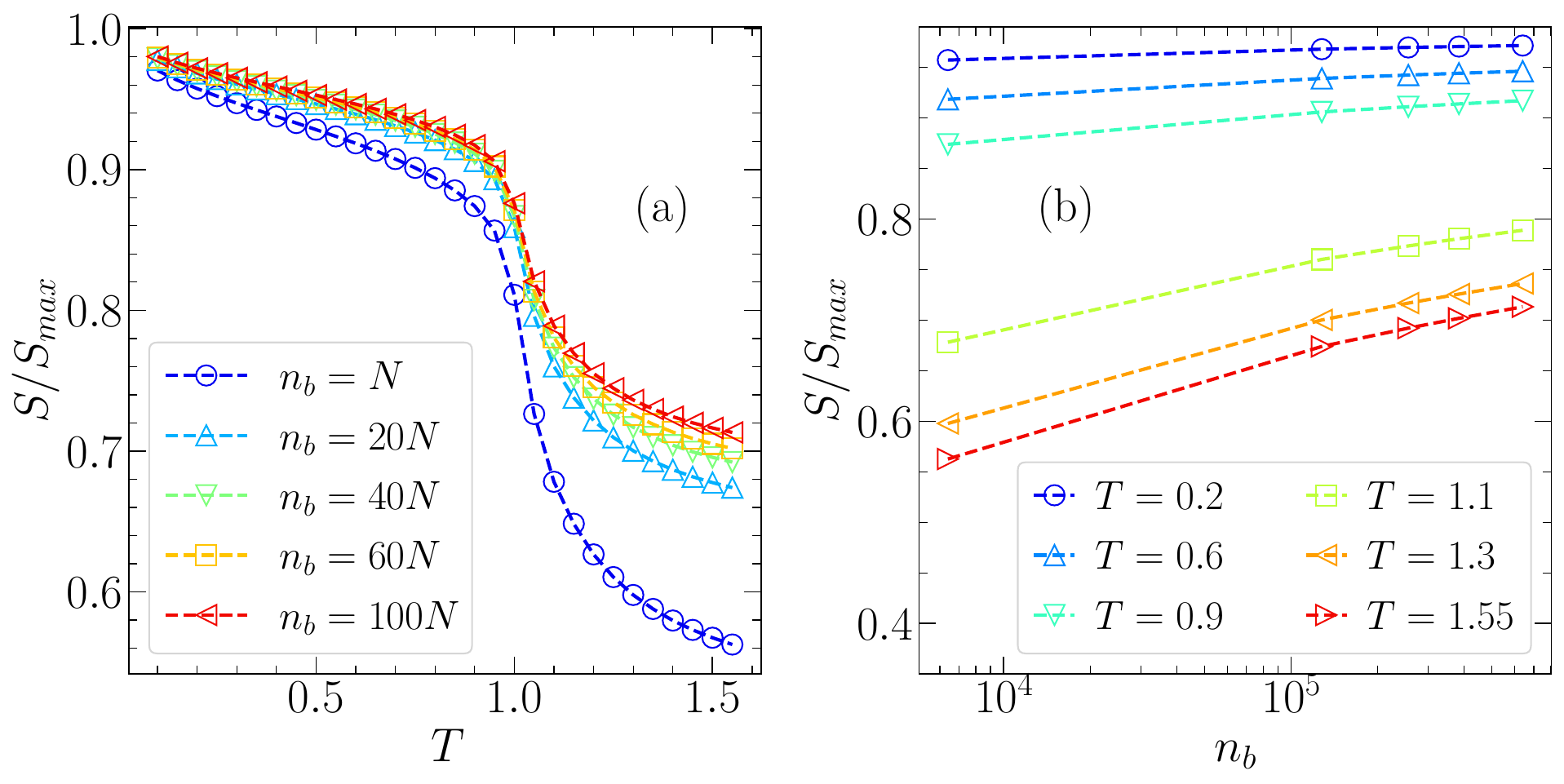}
\caption{(a) The participation entropy $S/S_\mathrm{max}$ verses $T$ for the $xy$ model with different $n_b$. (b) $S/S_\mathrm{max}$ as a function of $n_b$ for different temperatures.}
\label{fig:S_nb_xy}
\end{figure}

In the $XY$ model, the on-site spin configuration $\theta_i$ is continuous in the range $[0,2\pi]$, which eventually leads to a continuous distance $r$ between configurations. However, the participation entropy $S_r$ in Eq.~\eqref{eq:entropy} is defined for the discrete case. In the present work, for simplicity and consistency with other models, we keep this definition and compute $S_r$ by a discretization of $r$ with $n_b$ intervals. Then the probability $p(r)$ is counted in the interval $r\in [r-1/2n_b,r-1/2n_b]$, and the value $S_r$ depends on the choice of $n_b$. Fig.~\ref{fig:S_nb_xy}(a) shows the participation entropy as a function of temperature for different $n_b$, where $S/S_\mathrm{max}$ increases as $n_b$ increases with $S_\mathrm{max}=\log(n_b)$. However, all curves capture the anomaly that corresponds to the critical point. In Fig.~\ref{fig:S_nb_xy}, we explicitly show the $n_b$-dependence of $S_\mathrm{max}$, where the increment of $S_\mathrm{max}$ is quite slow. In the present, participation entropy is employed to extract the finite-size critical points, and the results in Fig.~\ref{fig:S_nb_xy} are enough to manifest the difference in different phases. In the main text, we use $n_b=100N$ for all results related to the participation entropy of the $XY$ model. 

Here is a brief explanation of why our calculation results for $S_r$ in XY model are related to $n_b$~.Let $\Delta$ be the width of each bin, obviously~ $\Delta=1/n_b$~here, and then $S_r$ calculated with ~$n_b$~bins can be expressed as
\begin{align}
    S_\Delta  &=- \sum_{i_b} p_{i_b} \ln(p_{i_b}) \notag\\
    &=- \sum_{i_b} P(r_{ib})\Delta \ln(P(r_{i_b})\Delta) \notag \\
    &=- \sum_{i_b} P(r_{ib})\Delta \ln(P(r_{i_b}))- \sum_{i_b} P(r_{ib})\Delta \ln(\Delta),
\end{align}
where $p_{i_b}$ represents the probability of $r$ appearing in the $i_b$-th interval. The probability density of $r_{i_b}$ equals to the average of the probability density in that interval, following the mean-value theorem. When $n_b$ is large enough, we can approximately assume that
\begin{align}
    S_\Delta&\approx -\int_{-\infty}^{\infty} P(r) \ln(P(r)) dr-\int_{-\infty}^{\infty} P(r)\ln(\Delta)dr \notag\\
    &=  -\int_{-\infty}^{\infty} P(r) \ln(P(r)) dr+\ln(n_b).
\end{align}
The first term in this equation represents Shannon's defined differential entropy, while the second term is the logarithm of $n_b$. When $T\to\infty$ and $N$ is large enough, the participation entropy of the $XY$ model reads
\begin{align}
    S_\Delta\approx\frac{1}{2}\ln(2\pi e/(12N))+\ln(n_b).
\end{align}

\section{Kagome Ising model}
\label{appendix:kim}

\begin{figure}[!b]
\includegraphics[width=\columnwidth]{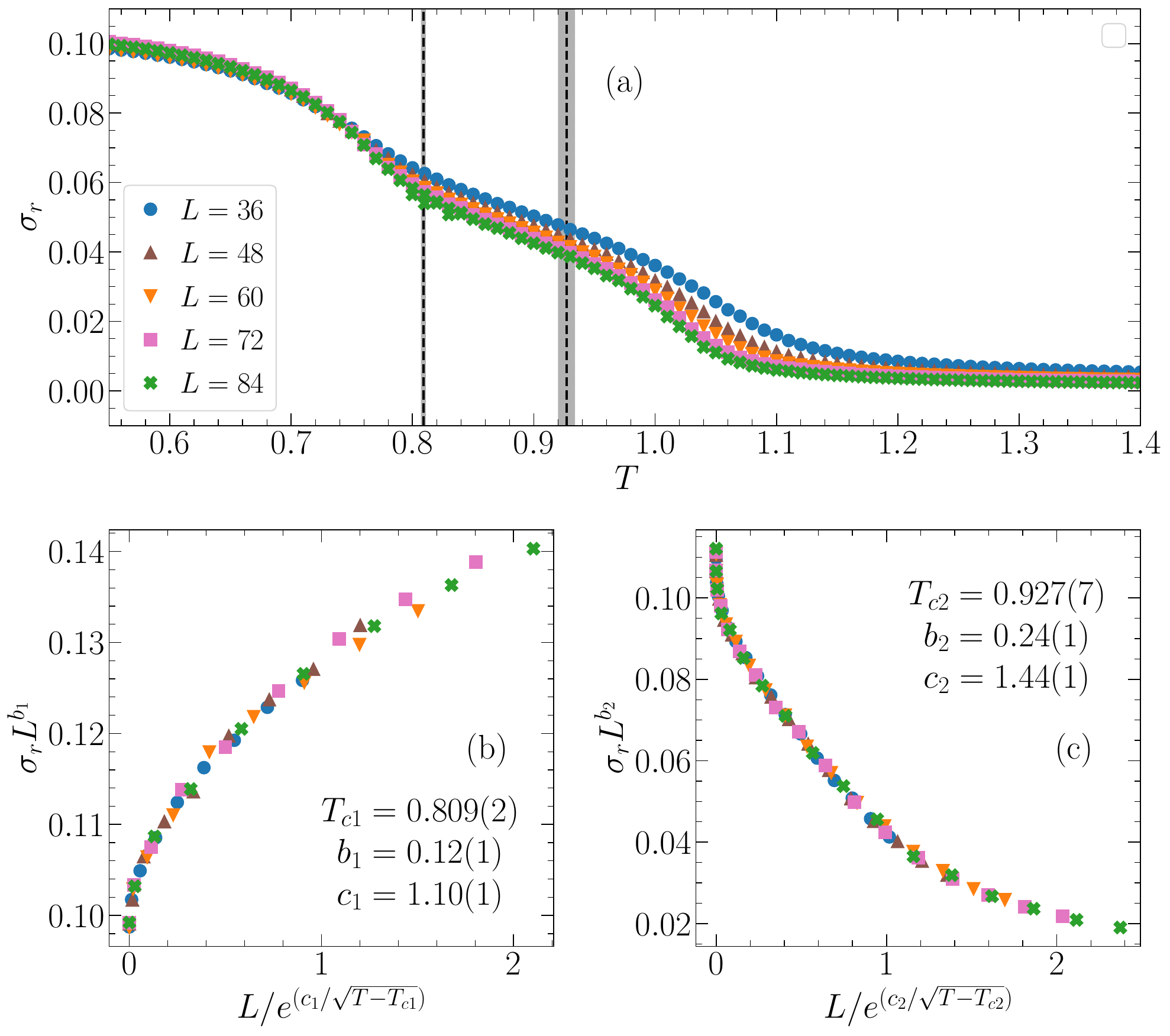}
\caption{(a) The standard deviation $\sigma_r$ with respect to temperature $T$ for the KIM with different $L$. The vertical dashed lines mark the critical points from the finite-size scaling of $\sigma_r$, with the shaded area depicting the error. (b) and (c) show the best data collapse of the scaling for $T_{c1}$ and $T_{c2}$, respectively. }
\label{fig:sigma_kim}
\end{figure}

The kagome spin ice can be described by the kagome Ising model (KIM) with antiferromagnetic nearest neighbor (NN) interactions and ferromagnetic next nearest neighbor (NNN) ones~\cite{chern2012, KZhao, Su2023}. The model Hamiltonian reads  
\begin{align}
H=J_1 \sum_{\langle ij\rangle} \sigma_i^z \sigma_j^z + J_2 \sum_{\langle\langle ij\rangle\rangle} \sigma_i^z \sigma_j^z,
\label{eq:ham}
\end{align}
where the sum $\langle ij\rangle$ ($\langle\langle ij \rangle\rangle$) runs over the NN (NNN) sites with $J_{1}>0$ ($J_{2}<0$). The KIM described by Eq.~\eqref{eq:ham} features the ground state of a six-state clock spin ice~\cite{chern2012,takagi1993}, which can be described by a complex order parameter
\begin{align}
M=\frac{1}{N}\sum\sigma_{i}{\rm exp}(\mathrm{i}\bm{Q}\cdot\bm{r}_{i})
\label{eq:op}
\end{align}
with $\bm{Q}=(4\pi/3,0)$. As temperature increases, the KIM is known to have two finite-temperature phase transitions, which belong to the same universality class as the six-state clock model~\cite{Wolf_kagome_1988}.

In this work, we choose $J_1=1$ and $J_2=-1/3$ as in Ref.~\cite{Su2023}. Although the order parameter is different, we take the same analysis based on distances between the sampled configurations, and the distance is defined the same way as for the Ising model on the square lattice in Eq.~\eqref{eq:distance_q2}. The finite-size scaling of $\sigma_r$ is displayed in Fig.~\ref{fig:sigma_kim}, where the critical points agree well with values contained from the scaling of order parameters~\cite{Su2023}.

\bibliography{main}

\end{document}